\documentclass[final,5p,times]{elsarticle}

\usepackage{graphicx}

\usepackage{amssymb}

\usepackage{xspace}
\usepackage{textcomp}

\usepackage{booktabs}
\usepackage{graphicx}
\usepackage{caption}
\usepackage{subcaption}
\usepackage{hyperref}
\usepackage{siunitx}

\newcommand{\Neq}{\ensuremath{\,\rm{n}_{\rm{eq}}/cm^2}}

\journal{Nuclear Instruments and Methods A}

\begin{document}
\begin{frontmatter}

\title{{\boldmath Irradiation study of a fully monolithic HV-CMOS pixel sensor design in AMS 180 nm}}

\author[a]{H.~Augustin}
\author[b]{N.~Berger}
\author[a]{S.~Dittmeier}
\author[a]{J.~Hammerich}
\author[a]{A.~Herkert}
\author[a]{L.~Huth\corref{cor1}}
\ead{huth@physi.uni-heidelberg.de}
\author[a]{D.~Immig}
\author[a]{J.~Kr\"oger}
\author[a]{F.~Meier}
\author[c]{I.~Peri\'c}
\author[a]{A.-K.~Perrevoort}
\author[a]{A.~Sch\"oning}
\author[b]{D.~vom Bruch\fnref{Parisfootnote}}
\author[a]{D.~Wiedner}

\cortext[cor1]{corresponding author}

\address[a]{Physikalisches Institut der Universit\"{a}t Heidelberg, INF 226, 69120 Heidelberg, Germany}
\address[b]{Institut f\"{u}r Kernphysik, Johann-Joachim-Becherweg 45, Johannes Gutenberg-Universit\"{a}t Mainz, \newline55128 Mainz, Germany}
\address[c]{Institut f\"{u}r Prozessdatenverarbeitung und Elektronik, KIT,\\ Hermann-von-Helmholtz-Platz 1, 76344 Eggenstein-Leopoldshafen, Germany}
\fntext[Parisfootnote]{Now at LPNHE, Sorbonne Universit\'e, Universit\'e Paris Diderot, CNRS/IN2P3, Paris, France}


\begin{abstract}
{\bf{H}}igh-{\bf{V}}oltage {\bf{M}}onolithic {\bf{A}}ctive {\bf{P}}ixel {\bf{S}}ensors (HV-MAPS) based on a 180~nm HV-CMOS process have been proposed to realize thin, fast and highly integrated pixel sensors. 
The MuPix7 prototype, fabricated in the commercial AMS H18 process, features a fully integrated on-chip readout, i.e.~hit-digitization, zero suppression and data serialization. MuPix7 is the first fully monolithic HV-CMOS pixel sensor that has been tested for the use in high irradiation environments like HL-LHC. 
We present results from laboratory and test beam measurements of MuPix7 prototypes irradiated with neutrons (up to  \SI{5.0e15}{\Neq}) and 24 GeV protons (up to \SI{7.8e15}{protons \per cm\squared})  and compare the performance with non-irradiated sensors. 
At sensor temperatures of about \SI{8}{\celsius} efficiencies of \SI{\ge  90}{\percent} at noise rates below 40 Hz per pixel are measured for fluences of up to \SI{1.5e15}{\Neq}.
A time resolution better than \SI{22}{ns}, expressed as Gaussian $\sigma$, is measured for all tested settings and sensors, even at the highest irradiation fluences. The data transmission at 1.25 Gbit/s and the on-chip PLL remain fully functional.
\end{abstract}

\end{frontmatter}

\section{Introduction}
 {\bf{H}}igh-{\bf{V}}oltage {\bf{M}}onolithic {\bf{A}}ctive {\bf{P}}ixel {\bf{S}}ensors (HV-MAPS) have been proposed \cite{Peric:2007zz} as an alternative to hybrid detectors. 
 High-Voltage CMOS processes allow for the implementation of strong electric drift fields
 in a thin active depletion region as small as \SI{10}{\micro m}.
 The main mechanism for charge collection  is therefore drift in contrast to standard MAPS where the charge is mainly collected by diffusion. 
 Due to the small depletion region for the standard low resistivity substrates (\SI[parse-numbers = false]{10-20}{\ohm cm}) and biasing from the top side, the inactive p-substrate can be removed allowing for \SI{50}{\micro m} thin sensors.  
 The small depletion region makes HV-CMOS designs also less sensitive to charge trapping caused by radiation damage of the bulk, which reduces the effective charge collection distance.
 
The AMS H18 process \cite{AMS} is a commercial HV-CMOS process providing high reliability at moderate production cost. It allows for the integration of analogue readout electronics into an active pixel matrix as well as the implementation of digital electronics as seen in figure~\ref{fig:hvmaps}. 
In this paper the performance of a fully monolithic\footnote{I.e. signal amplification, digitization, zero supression, time stamp sampling, readout state-machine and data serialization is implemented on chip.} HV-MAPS prototype after proton and neutron irradiation is studied for the first time.

For another sensor with a similar pixel cell design and
hybrid readout (CCPDv4~\cite{ccpdv4}), it was shown that the analogue pixel performance  remains high
after particle fluences of up to \SI{5.0e15}{1~MeV \Neq} \cite{ccpdv4Irr}. 
Additional studies with prototypes of {\bf{D}}epleted {\bf{M}}onolithic {\bf{A}}ctive {\bf{P}}ixel {\bf{S}}ensor
(DMAPS) based on a modified \SI{180}{nm} CMOS process from TowerJazz~\cite{towerJazz} have
also shown high efficiencies after irradition with up to \SI{1e15}{\Neq}.

The MuPix HV-MAPS prototypes studied here have been developed for the Mu3e experiment \citep{RP} using the IBM/AMS H18 process. 
To test the radiation tolerance, ten MuPix7 prototype sensors were irradiated without bias at room temperature with {24 GeV protons at the CERN PS \cite{PS} and with neutrons at the TRIGA reactor at the  Jo\v{z}ef Stefan Institute~\cite{Ljubljana} in Ljubljana, with particle fluences up to  \SI{7.8e15}{protons/cm\squared} and \SI{5.0e15}{neutrons/cm\squared}, respectively.

\section{MuPix7 prototype}
The fully monolithic MuPix7 prototype \citep{Augustin2017194} is the first HV-MAPS integrating all functionalities required for a data driven readout, realized on a \SI[parse-numbers = false]{10-20}{\ohm cm} substrate.
It consists of a \SI[parse-numbers=false]{32\times40}{} pixel matrix with a pixel size of \SI[parse-numbers = false]{103\times80}{\micro\meter\tothe{2}}, resulting in an active size of \SI{10.5}{mm\squared} and a total size of \SI{15.58}{mm\squared}. 
Each pixel cell integrates a charge sensitive amplifier with a source follower driving the signal to the periphery, see figure~\ref{fig:mupixCircuit}. 
In the periphery, the analogue signal is amplified in a second stage and discriminated against a threshold which can be fine-adjusted for each pixel.
The pixel address and an \SI{8}{bit} time stamp generated at \SI{62.5}{MHz} are stored and read out by an on-chip state machine. 
The zero suppressed data are \SI{8}{bit}/\SI{10}{bit} encoded, serialized and transmitted at a rate of \SI{1.25}{Gbit/s} over a SCSI II cable with a length of \SI{1.8}{m}. 
The MuPix7 prototype does not contain additional measures like ring transistors or triple redundant logic to increase the radiation tolerance of the sensor.
The MuPix7 has a set of configuration DACs to adjust the response of the amplifier, the line driver strength, the state machine frequency, etc. 
These global MuPix configuration settings have been optimized once for the non irradiated sensors,  corresponding to a power consumption of \SI{300}{mW\per cm\squared} and are used without changes for all irradiated MuPix prototypes.

Previous studies have shown that non-irradiated MuPix7 sensors have an efficiency of above
\SI{99}{\percent} at per-pixel noise rates below \SI{20}{Hz}  and a depletion
voltage of \SI{-85}{V} \citep{Augustin2017194,mp7,mp7timing}. 
The time resolution $\sigma$ was measured to be below \SI{14.2}{ns}.
When using an
Fe-55 source approximately 1600 primary electrons are expected for an non-irradiated
MuPix7 with an average signal of \SI{200}{mV}  \cite{KLS}.
   
Four dedicated columns of the MuPix7 matrix implement single stage amplification and are not considered in the presented measurements.
\begin{figure}[hbtp]
	\centering
		\includegraphics[width=.8\columnwidth,keepaspectratio]{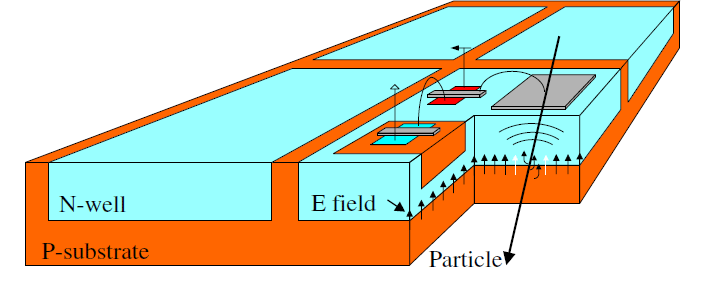}
	\caption{\label{fig:hvmaps} Schematic view of the working principle of HV-MAPS.
		\cite{Peric:2007zz}. }
\end{figure}

 \begin{figure}[hptb!]
\centering
\includegraphics[width=\columnwidth,keepaspectratio]{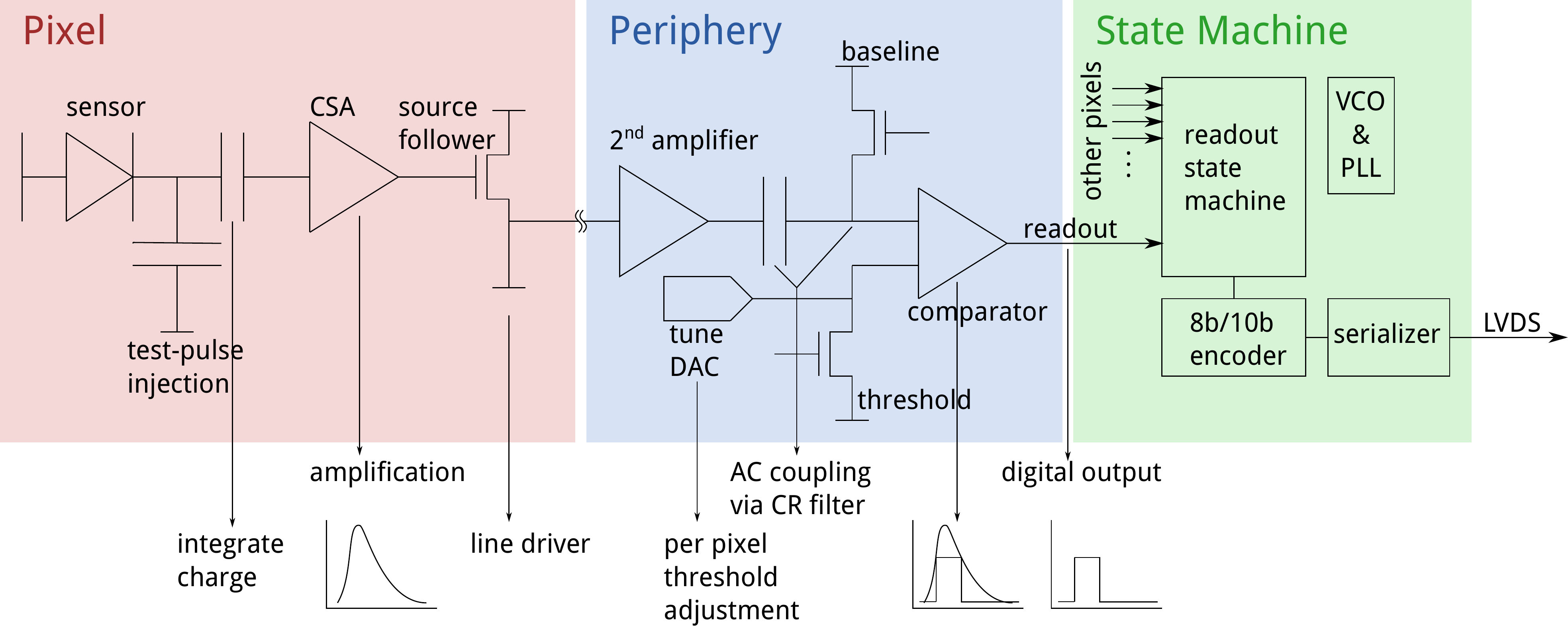}
\caption{\label{fig:mupixCircuit} Sketch of the MuPix7 readout circuitry.}
\end{figure}

\subsection*{Irradiated samples}
MuPix7 sensors have been irradiated with protons and neutrons according to
table~\ref{tab:sensors}. During irradiation, the sensors have not been biased
and not cooled. The prototypes had not been characterized beforehand.  The
samples have sensor thicknesses of \SI{60}{\micro m} and \SI{75}{\micro
  m}. The studies have been performed after one year of annealing at room
temperature and therefore complete annealing of the ionizing damage is
expected.
The non-ionizing damage responsible for charge trapping and bulk damage,
however, is expected to be left unchanged and is the
primary focus of this study.
\begin{table*}[tbp]
\centering
\begin{tabular}{ r  |l l l l| l l l l}
\toprule
Sensor ID & P00 & P814& P1515 & P7815 & N00 & N514 & N115 & N515 \\ \midrule
Facility & - & PS & PS & PS & - & TRIGA & TRIGA & TRIGA  \\
NIEL Fluence {[1 MeV \Neq]} & 0 &\SI{4.8e14}{}&\SI{9e14}{}&\SI{4.7e15}{}&0 &\SI{5e14}{}&\SI{1e15}{} & \SI{5e15}{}\\
Proton Fluence {[24 GeV/c $\,$\SI{}{p/cm\squared}]} & 0 &\SI{0.8e15}{} &\SI{1.5e15}{} & \SI{7.8e15}{} &-&-&-&-\\
\bottomrule

\end{tabular}
\caption{\label{tab:sensors}  List of irradiated sensors. The quoted fluences
  are averaged over the sensor area. A hardness factor of
  \SI{0.6}{\rm{n}_{\rm{eq}}/p} \cite{PS_NIEL} is used to calculate the NIEL fluences for the proton irradiated samples. The sample P814 was not characterized during the test beam.}
\end{table*}
\section{Setup} \label{sec:setup}
The proton irradiated sensors are directly glued and wire-bonded to a printed-circuit-board (PCB).
The neutron irradiated sensors are glued and wire-bonded to a ceramic carrier, which is connected to the same PCB type via a socket\footnote{Due to limited availability of PCBs only the proton irradiated sensors were directly bonded to PCBs.}. 
The PCB provides stable and ripple-free power, filters the high voltage and converts the differential slow control signals to single ended signals required by the MuPix7. 
Baseline and threshold reference for the sensor are generated on the PCB where also test pulses to mimic signals can be generated.

The test beam setup is sketched in figure~\ref{fig:cooling}.
The sensors under test are actively cooled to reduce leakage currents and noise
while also preventing thermal runaway. 
The proton irradiated sensors are cooled by cold nitrogen gas flowing over the backside of the sensor. 
The neutron irradiated sensors are cooled using a Peltier element, connected with an aluminum chuck. 
The Peltier element in turn is cooled by cold nitrogen gas.
Two thermal baths are used to cool the nitrogen gas to  \SI{-20}{\celsius} for both setups.
The cooling power of the setup is controlled by the applied gas flow.
A combined nitrogen flow of \SI{2.5}{m\cubed\per\hour} is used to cool the two sensors under test.
The temperature of both devices-under-test (DUTs), as well as the humidity in the box, are continuously  monitored to guarantee safe and constant operation conditions. 
Two reference tracking telescopes \cite{TWEPP2017}, consisting of three non-irradiated, uncooled MuPix7 sensors each, are used to measure efficiencies.
 \begin{figure}[tbp]
\centering
\includegraphics[width=.9\columnwidth,keepaspectratio]{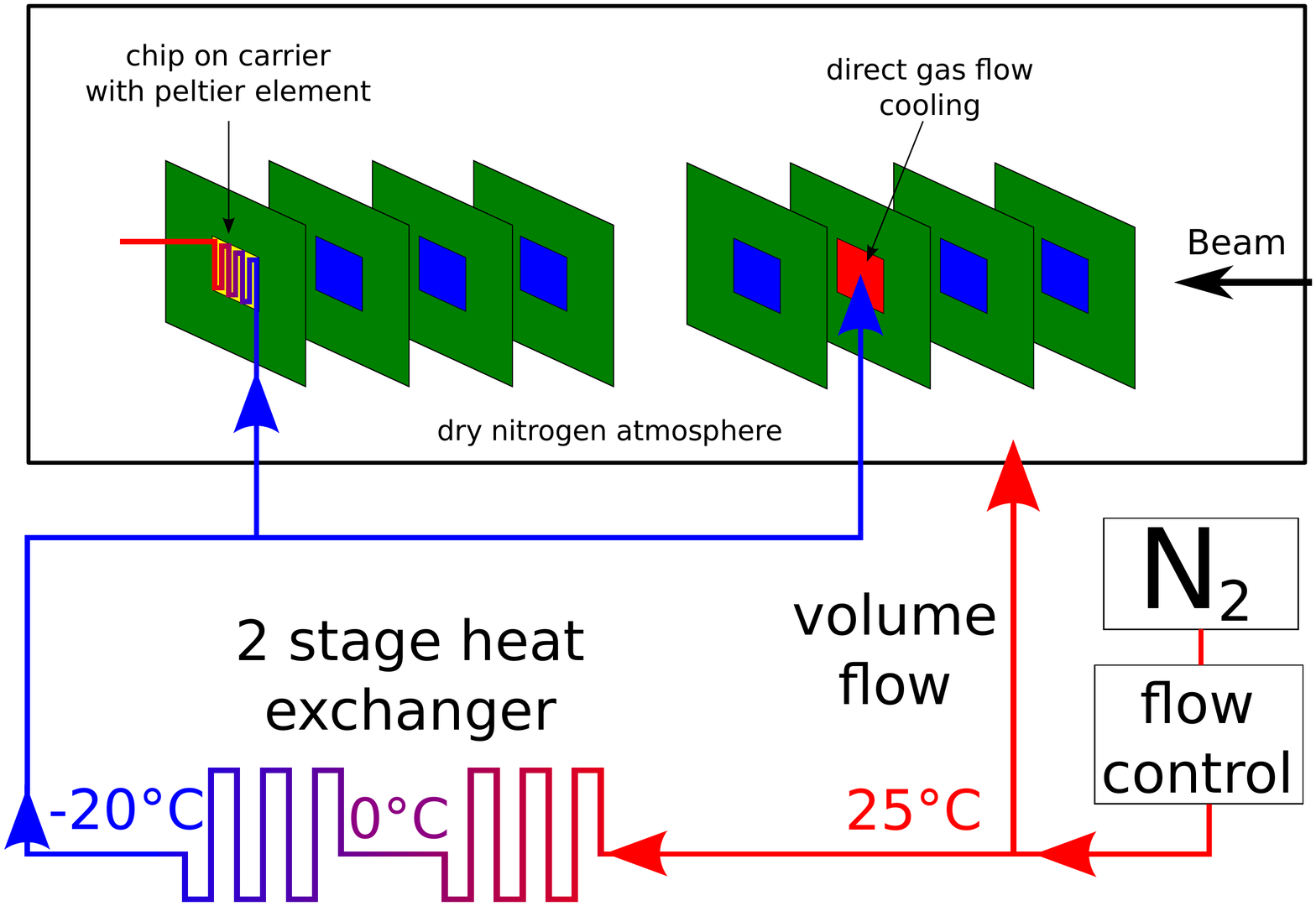}
\caption{\label{fig:cooling} Sketch of the cooling setup in test beam. The red squares
  represent the devices under test. The blue squares represent reference sensors used for performance studies.}
\end{figure}

\subsection*{Temperature calibration}

The temperature of the proton irradiated MuPix is monitored by measuring the gas temperature close to the MuPix. 
The temperature of the neutron irradiated prototypes is monitored by measuring the cooling chuck temperature. 
For both devices the MuPix temperature $\text{T}_{\text{MuPix}}$ is calculated from the
measured temperatures $\text{T}_{\text{meas}}$ by applying a correction 
obtained from an IR-camera\footnote{The IR-camera in turn had been calibrated
  using a Pt1000 \cite{PT1000} glued on a heatable reference silicon surface.}. 
The correction $\text{T}_{\text{MuPix}}-\text{T}_{\text{meas}}$ is about \SI[parse-numbers = false]{5 (-6)}{\celsius} for the proton (neutron) irradiated prototypes. 
The uncertainty on the absolute temperature is dominated by the reproducibility of the thermal coupling between MuPix and temperature sensor and estimated to be  $\pm$~\SI{2}{\celsius}.
The relative uncertainty of the measured temperature over time is small and  mainly given by the temperature sensor's uncertainty of \SI{0.7}{\celsius}. 
The temperature is stable on the  $\pm$~\SI{1}{\celsius} level for measurement periods of \SI{15}{hours}, see figure~\ref{fig:T_stab}. 
The MuPix temperature for both, proton and neutron irradiated, sensors is approximately \SI{8}{\celsius}.

\begin{figure}[tbp]
			\includegraphics[width=\columnwidth,keepaspectratio]{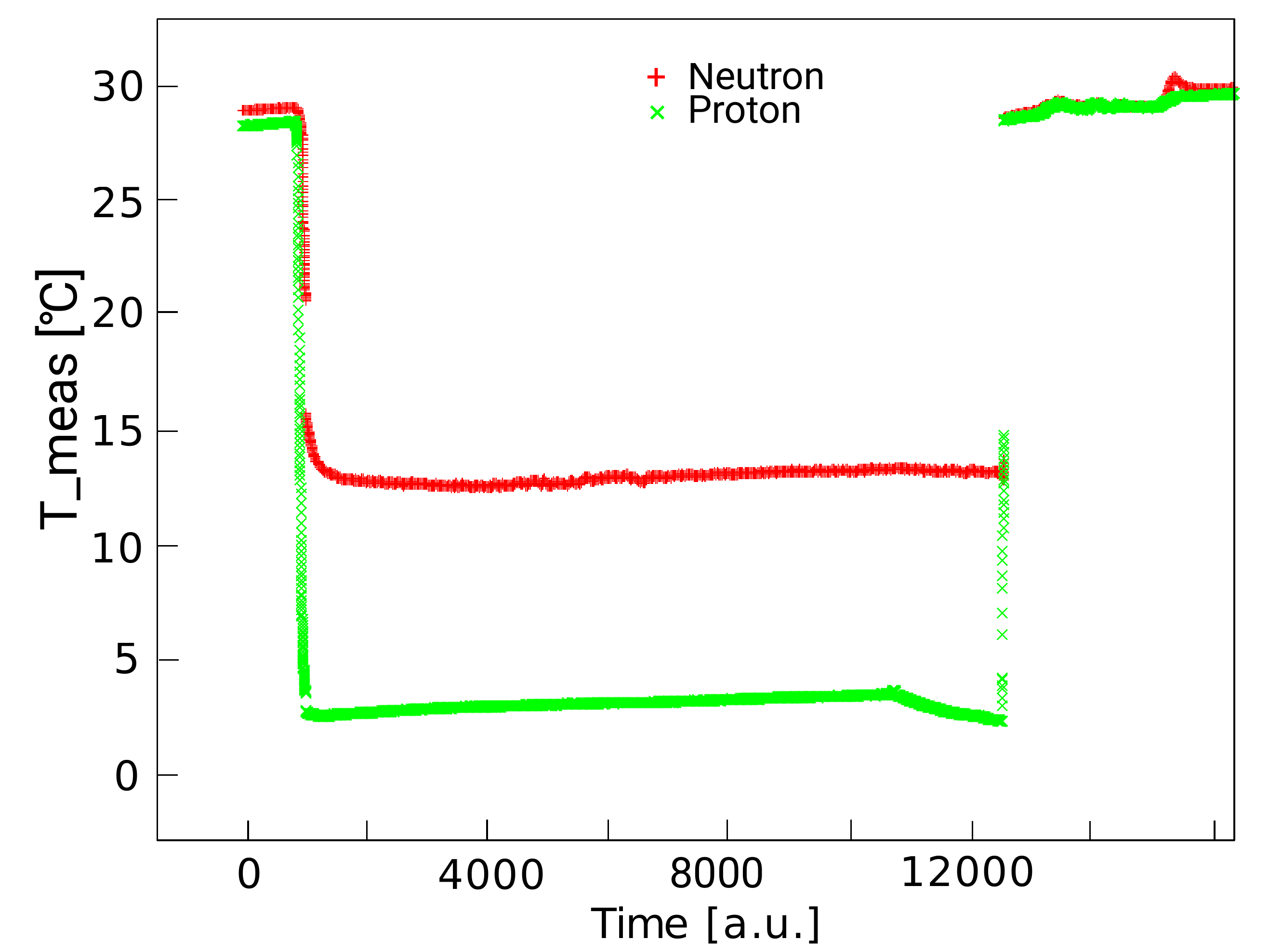}
		\caption{\label{fig:T_stab} Monitored temperature
                  $\text{T}_{\text{meas}} $ of the gas flow close to the
                  proton irradiated MuPix ({\sl green}) and the cooling chuck
                  of the neutron irradiated MuPix ({\sl red}) over a 15 hours period.}
\end{figure}

\section{Characterization of irradiated sensors}
\subsection{Laboratory results}
All irradiated MuPix7 are fully operational after irradiation: the PLL can be locked to an external 125 MHz reference oscillator and the serial data output runs without \SI{8}{bit}/\SI{10}{bit} errors. All hit addresses and time stamps transmitted to the FPGA are checked to be logically correct, thus indicating a fully functional readout state machine and serializer.

\subsection*{Leakage currents}
\label{sec:leakage}

\begin{figure}[tbp]
		\begin{subfigure}{\columnwidth}
		\includegraphics[width=\columnwidth,keepaspectratio]{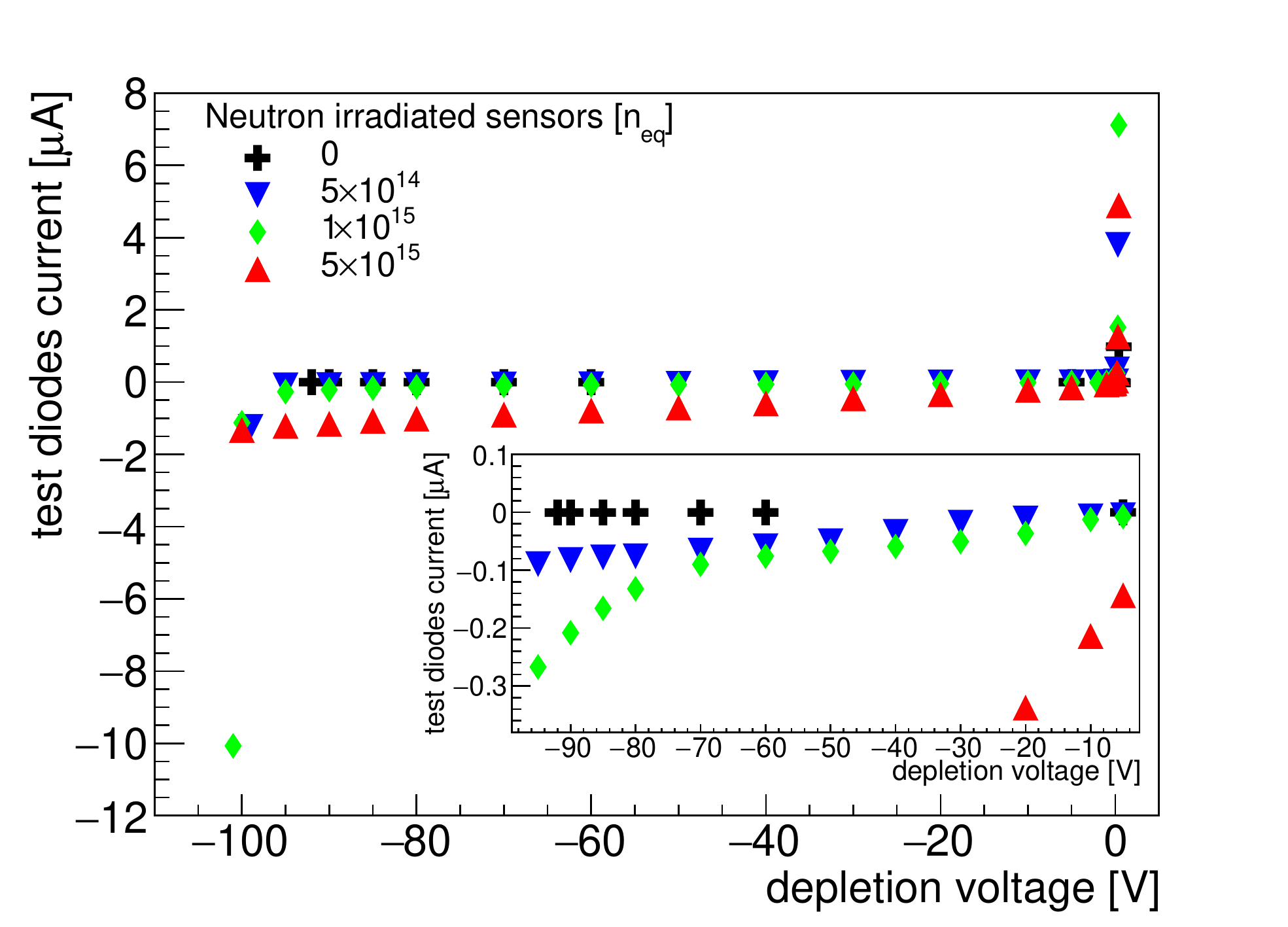}
		\caption{\label{fig:IV_n} Neutron irradiated samples.}
		\end{subfigure}
		\begin{subfigure}{\columnwidth}
		\includegraphics[width=\columnwidth,keepaspectratio]{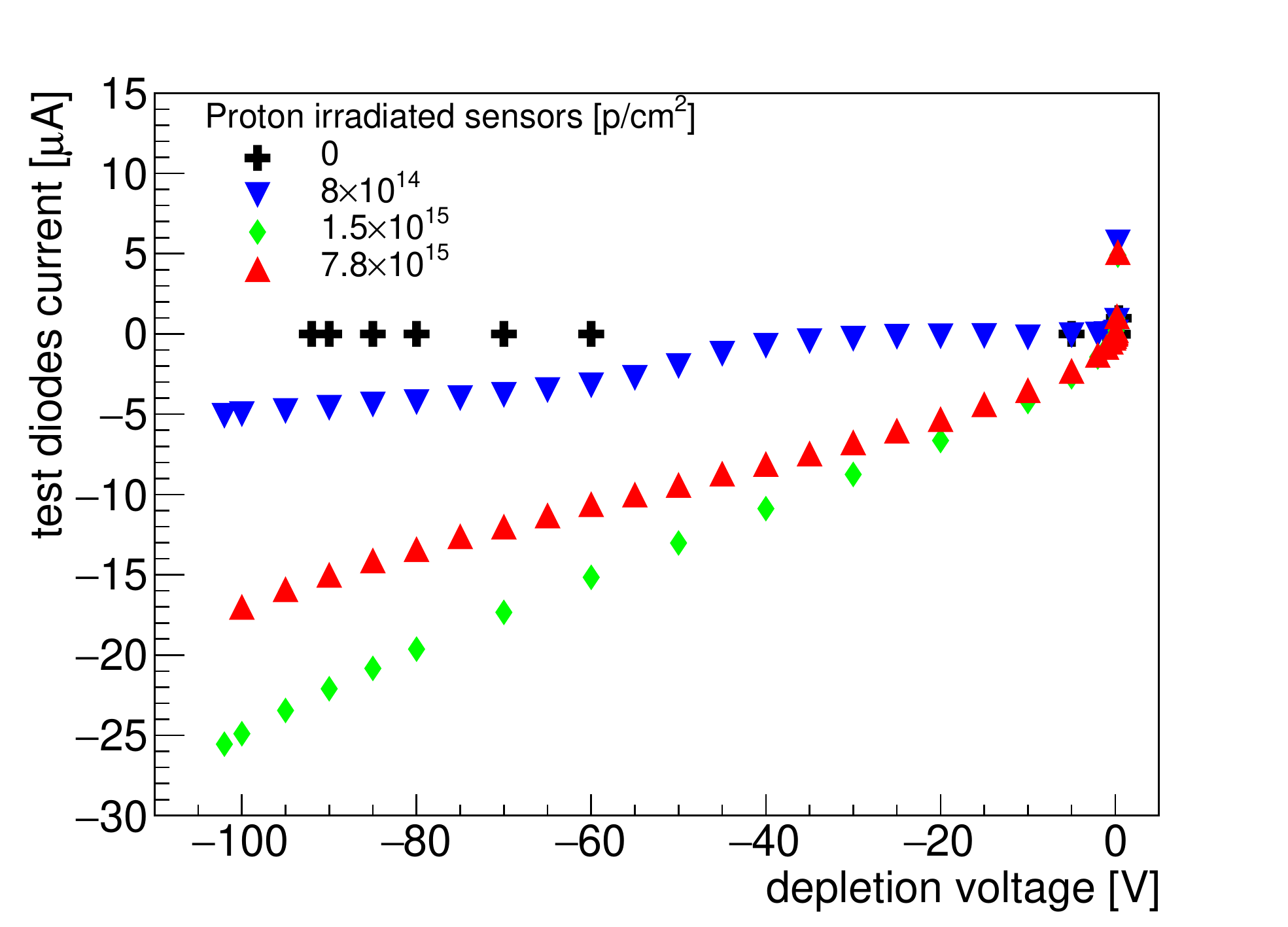}
		\caption{\label{fig:IV_p} Proton irradiated samples.}
		\end{subfigure}
	
	\caption{\label{fig:IV} Leakage currents of 32 test diodes with a total area of \SI{9.9e3}{\micro m \squared} as function of
          the depletion voltage for different irradiation levels.
          The measurements
          are performed at room temperature without cooling at $\text{T}_{\text{MuPix}} \approx$\SI{24}{\celsius}. }	
	\end{figure}
	
The leakage currents of non-irradiated and irradiated sensors are measured as a function of the depletion voltage. 
Figures \ref{fig:IV_n} and \ref{fig:IV_p} show the  current-voltage
characteristic (IV curve) for the neutron- and proton-irradiated sensors at room temperature when only depletion voltage is applied to a set of dedicated test diodes. 
As expected the measured leakage current significantly increases with the particle fluence and applied voltage. 
However, the absolute  leakage current increase for the proton irradiated
sensors is about twenty times higher than for the neutron irradiated sensors
for similar fluences. 
A possible explanation is that the hardness factor for \SI{24}{GeV} protons is significantly larger than expected
 for low-ohmic silicon wafers and the AMS-H18 process\footnote{Similar effects, indicating an enhanced acceptor removal in low ohmic substrates, have been reported by other groups and are under investigation.}.
It is also a bit surprising that the \SI{7.8e15}{protons/cm\squared} sample
shows less leakage currents than the  \SI{1.5e15}{protons/cm\squared} sample.
From edge-TCT measurements~\cite{h18_depletion} it is known that the depletion thickness has a maximum for a fluence
of about \SI{1.5e15}{protons/cm\squared} which possibly could explain the higher leakage currents.

The diode breakdown voltage is below \SI{-100}{V} for all irradiated
samples\footnote{The exact value of the breakdown voltage was not measured to
  minimize the risk of damage.}
and is significantly lower than the non-irradiated sensors, which have a typical breakdown voltage of about \SI{-90}{V}. This is consistent with an increased depletion thickness after irradiation \citep{h18_depletion}.

	\begin{figure}[htbp]
	\includegraphics[width=\columnwidth,keepaspectratio]{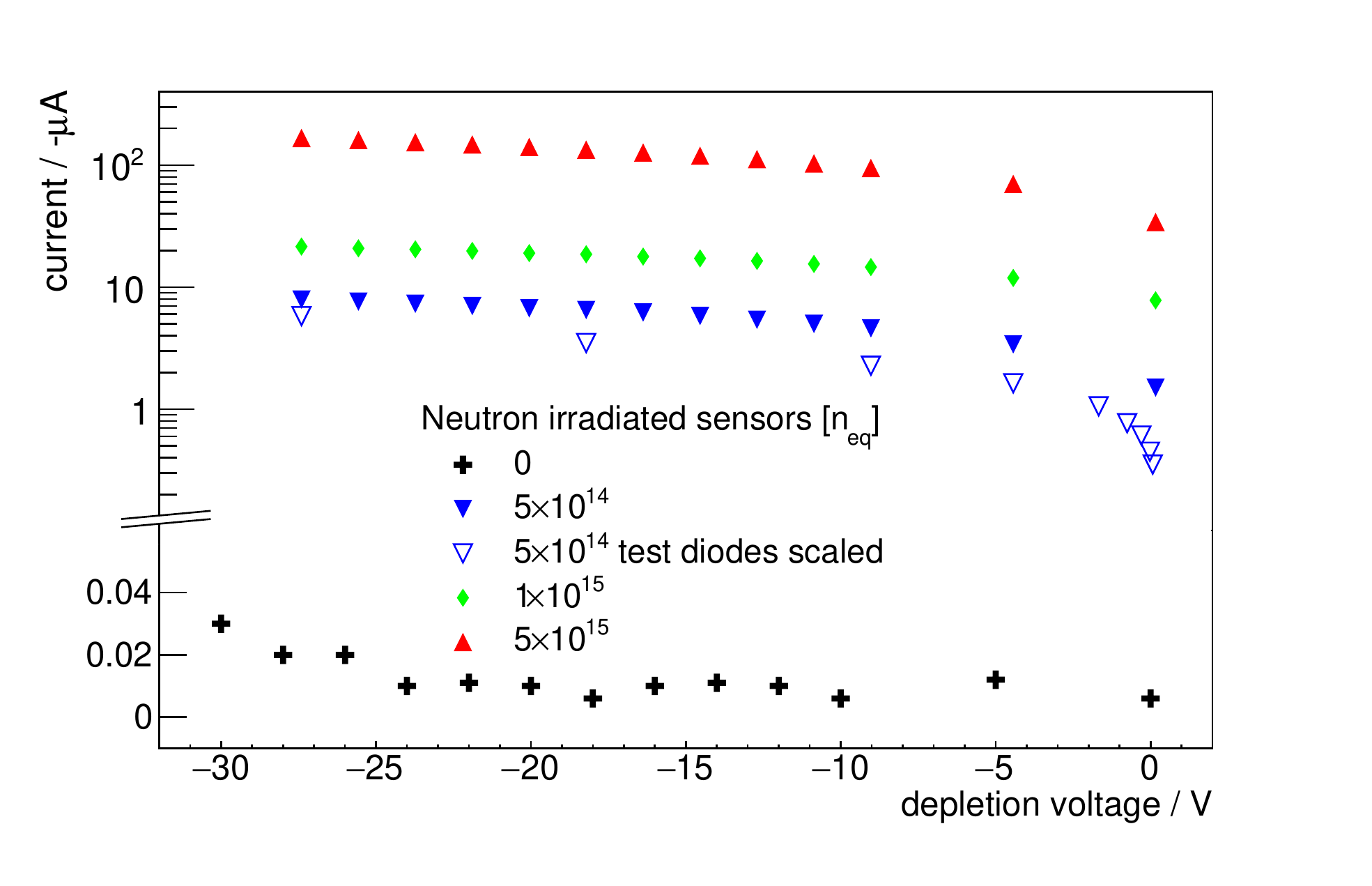}
		\caption{\label{fig:IV_active} Leakage currents of the
                  complete 32x40 pixel matrix for neutron irradiated sensors
                  in full operation for a sensor temperature
                  T$_{\text{MuPix}}$ $\approx$ \SI{40}{\celsius}. Each pixel
                  encloses 9 diodes.  The currents for the non-irradiated
                  sensors are below 40 nA. The guard rings are shorted to the bias. The test diode current from figure~\ref{fig:IV_n} is scaled by the diode size increase and shown for comparison.}
	\end{figure}
 
The leakage currents of all irradiated sensors in full operation are shown in figure~\ref{fig:IV_active} as a function of the depletion voltage. 
The slightly larger current compared to the measurements of the test diodes in figure~\ref{fig:IV_n} can be explained by the higher temperature of the sensor in full operation, see figure~\ref{fig:mupix_ir}.
Without active cooling it is not possible to operate the sensors with depletion voltages below \SI{-30}{V} due to thermal runaway.
With cooling,  for T$_{\text{MuPix}} \lesssim$~\SI{15}{\celsius}, the MuPix
can be operated up to a depletion voltage of \SI{-85}{V}.
The following measurements are obtained with the active cooling system
described in section~\ref{sec:setup} at a MuPix temperature of  about \SI{8}{\celsius}. 

\begin{figure}[htb]
	\centering
	\includegraphics[width=.8\columnwidth,keepaspectratio]{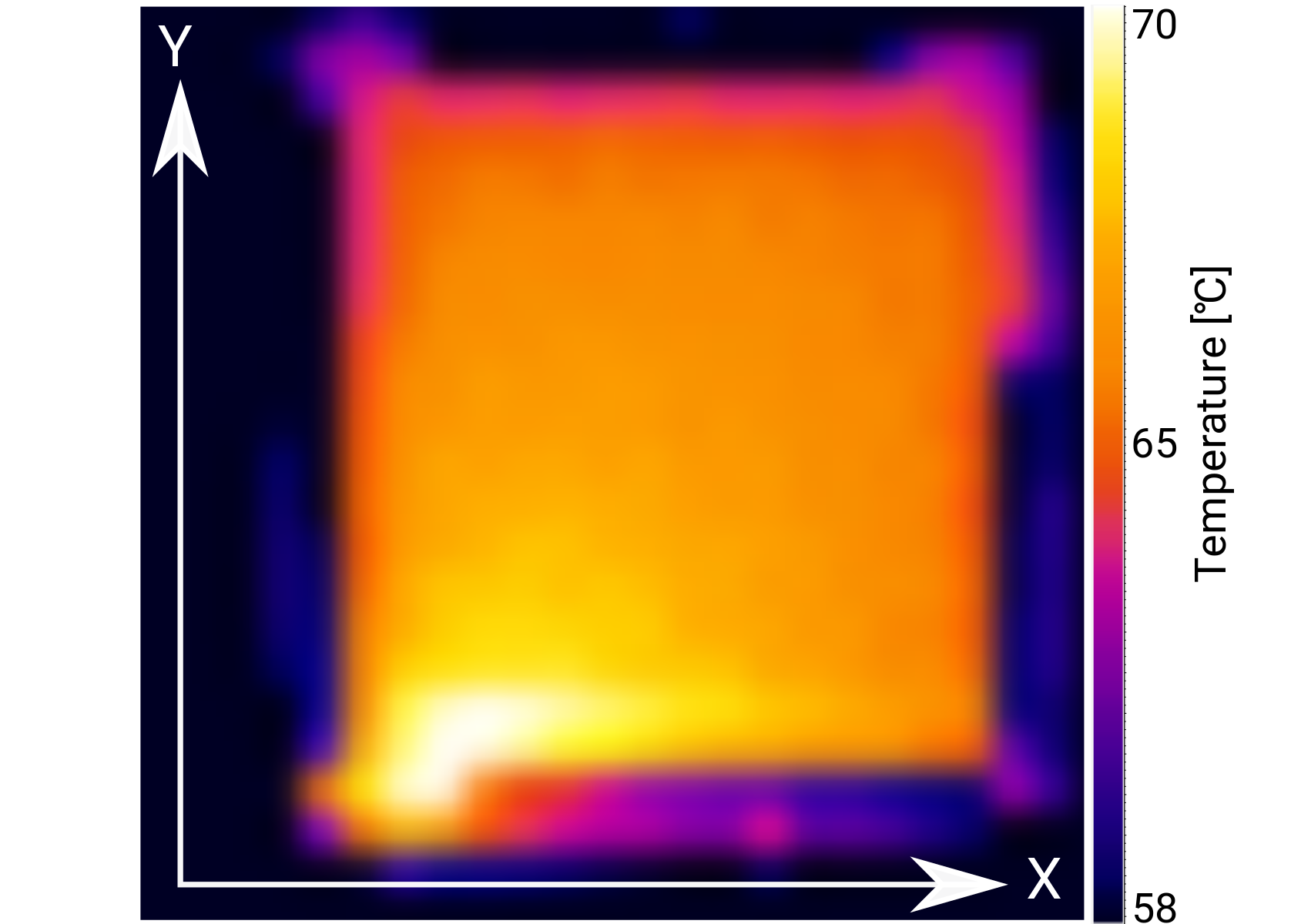}
	\caption{\label{fig:mupix_ir}IR camera picture of the
          \SI{1.5e15}{\Neq} irradiated MuPix in full operation. The fast
          digital logic of the state machine and the serializer is located in
          the bottom left part of the chip which shows with 
\SI{85}{\celsius} (\SI{70}{\celsius} uncalibrated) the highest temperature.}
\end{figure}
\noindent

\subsection*{Pixel tuning}
In MuPix7, pixel-to-pixel variations can be corrected for with a \SI[parse-numbers = false]{4-}{bit} tuning DAC (TDAC), by adjusting the discriminator threshold. 
A global DAC (VPDAC) selects from the externally provided threshold reference the tuning range covered by the TDACs.
Higher TDAC values correspond to higher effective thresholds, i.e. the
separation between threshold and baseline is increased.
The pixel response adjustment is based on a noise measurement and
an automated software calibration scheme is used. The threshold reference and the VPDAC value are selected such that all pixels are below a certain noise rate at the maximal TDAC value. 
Afterwards, each individual  TDAC value is adjusted such that the noise rate of every pixel is just below the target noise rate (\SI{1}{Hz} without beam). 
The thresholds obtained by the automated tuning routine strongly vary between sensors and depend on the applied depletion voltage and the irradiation fluence.
A summary of the tested sensors, the corresponding tune thresholds and VPDAC values is given in table~\ref{tab:tunings}.
For the proton irradiated samples, the described tuning procedure does not work properly for the complete matrix because of too large pixel-to-pixel variations caused by the inhomogeneities in the irradiation process (see below).  
Therefore, up to 5\% of all pixels\footnote{Pixels cannot be masked in the MuPix7.} are excluded in the presented analysis.
\begin{table*}[htbp]
	\centering
	\begin{tabular}{rclc|rclc}
		\toprule
		ID & HV [-V]& threshold&VPDAC &    ID & HV [-V]& threshold&VPDAC \\
		& & reference [mV]& &    & & reference  [mV]& \\
		\midrule
		P00 & 40 & 735&38	& N00 	 & 40 & 700 & 19\\
		& 60 & 725&21	&   	 & 60 & 718 & 20\\
		& 70 & 740&21	&   	 & 70 & 725 & 19\\
		& 85 & 734&20	&   	 & 85 & 725 & 19\\
		P1515 & 60 & 700&19	& N514	 & 60 & 740 & 18\\
		& 70 & 680&19	& N115	 & 40 & 711 & 23\\
		& 85 & 568&20	& 	 	 & 60 & 711 & 22\\
		P7815 & 40 & 770&26	& 	 	 & 70 & 740 & 23\\
		& 60 & 755&26	& 	 	 & 85 & 730 & 24\\
		& 70 & 760&28	& N515	 & 40 & 675 & 22\\
		& 75 & 740&27	& 	 	 & 60 & 675 & 21\\
		&    &    &	&	 	 & 70 & 690 & 25\\
		&    &    &    & 		 & 85 & 690 & 25\\
		\bottomrule
	\end{tabular}
	\caption{HV and configuration parameters used for chip
          characterization. The target noise rate for the pixel tuning is
          \SI{1}{Hz\per pixel} at the given threshold references. The beam is switched off during tuning. The baseline is externally applied and set to about \SI{800}{mV} for all settings. 100 mV signal height correspond to approximately 800 primary electrons \cite{KLS}.}
	\label{tab:tunings}
\end{table*}

The TDAC values and their distribution over the chip after tuning are shown in figure~\ref{fig:tuning} for three sensors. 
The TDAC map of the neutron irradiated MuPix (figure \ref{fig:ntune}) looks
similar to the TDAC map for a non-irradiated MuPix (figure~\ref{fig:tune}).
For the proton irradiated sensors (figure~\ref{fig:ptune})
a non-uniform distribution of the TDACs is obtained  which we attribute to a
non-uniform irradiation beam profile.
The observed non-uniform radiation damage causes position dependent depletion and varying noise levels \cite{ChargeCollection}.
The distribution of the TDAC values (figure \ref{fig:ptune}) has a shoulder towards lower
values which originate from the bottom right part of the TDAC map 
where the irradiation beam center was probably located \footnote{Dosimetry
  results of the facility suggest a 10\% dose variation over the sensor.}.
\begin{figure}[tbp]
\centering
\begin{subfigure}[t]{\columnwidth}
	\includegraphics[width=.5\columnwidth,keepaspectratio]{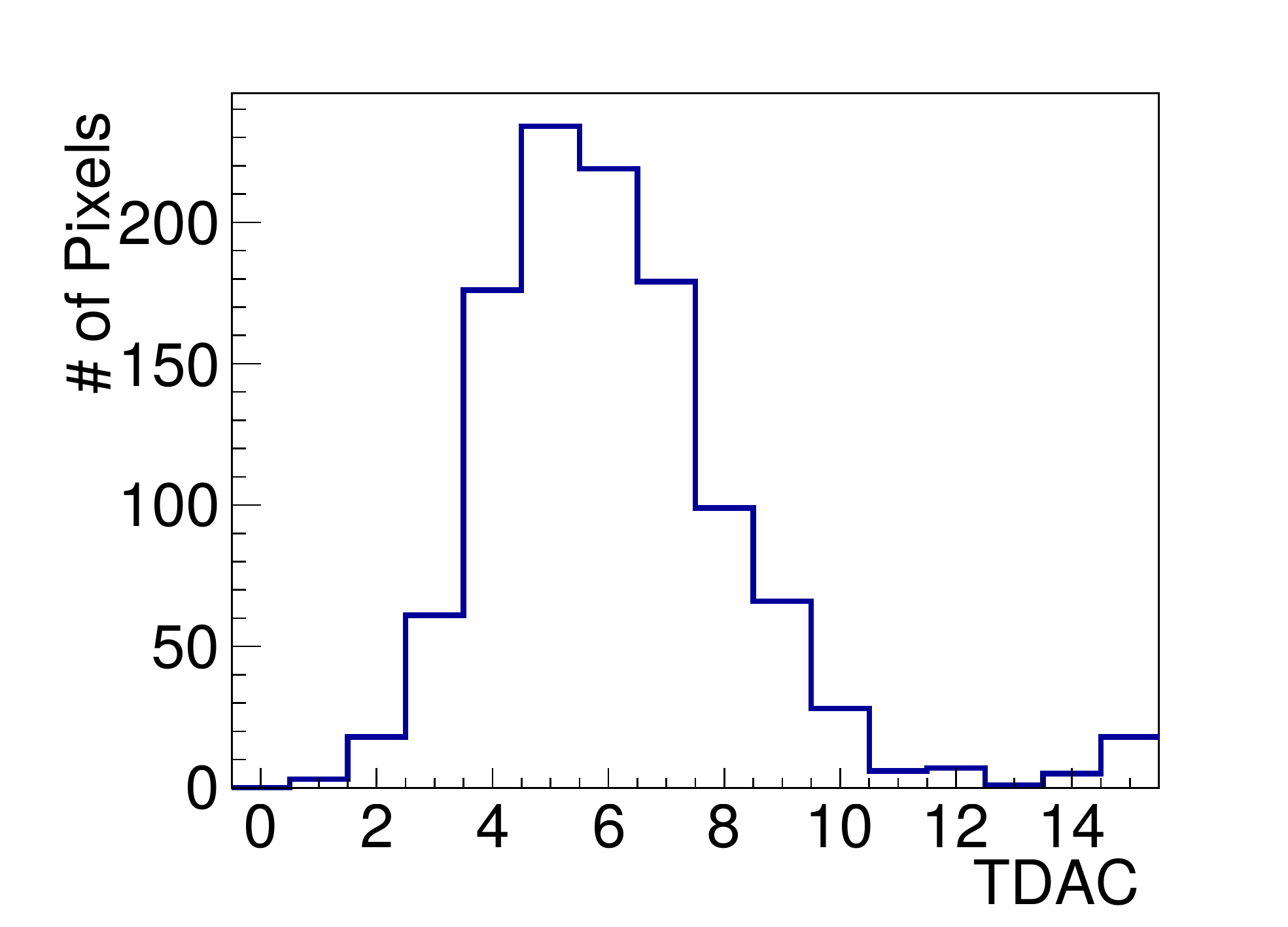}\hspace*{-0.3cm}
	\includegraphics[width=.5\columnwidth,keepaspectratio]{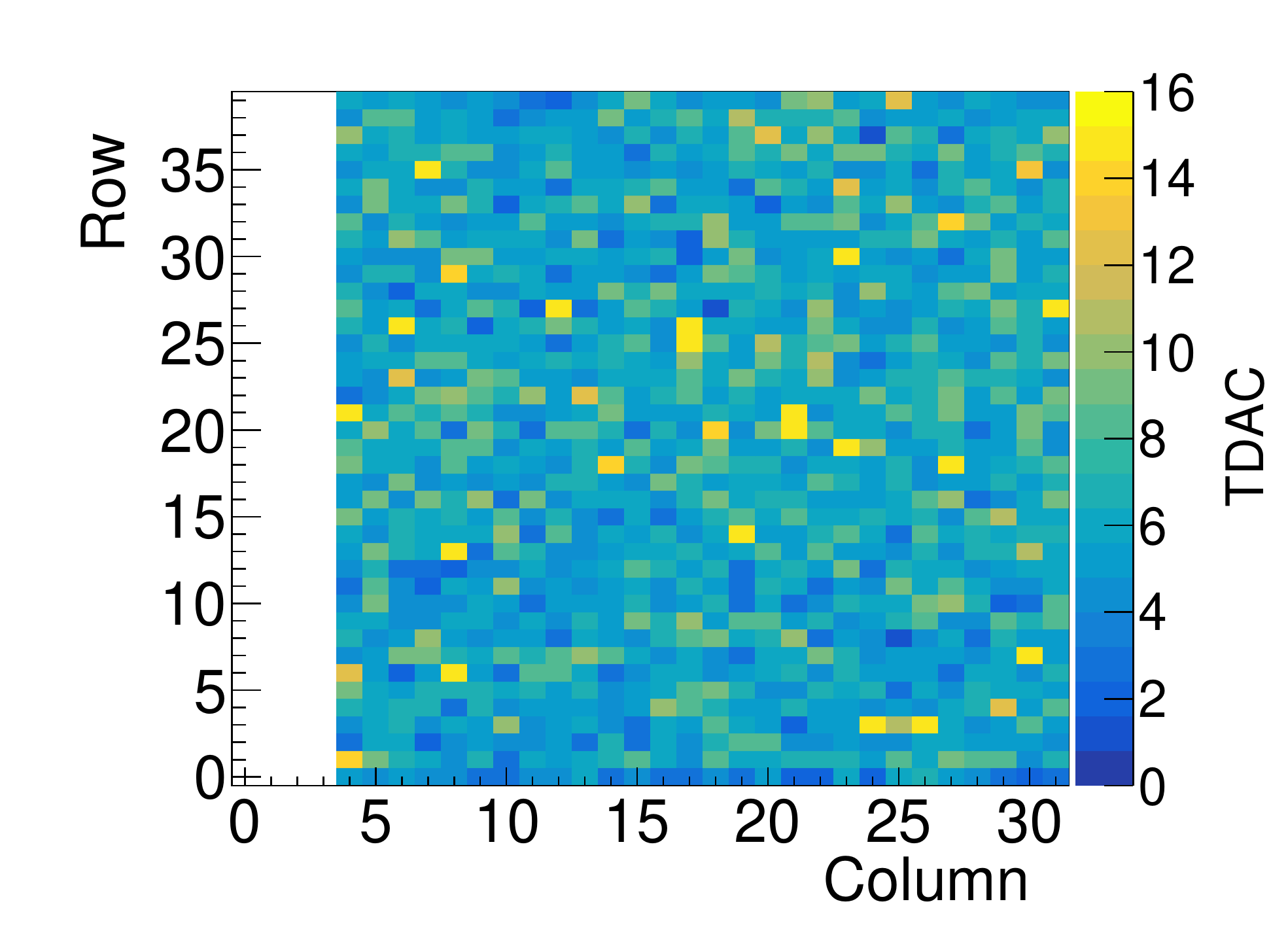}
	\caption{\label{fig:tune}non-irradiated (ID~P00)}
\end{subfigure}
    \begin{subfigure}[t]{\columnwidth}
    \includegraphics[width=.5\columnwidth,keepaspectratio]{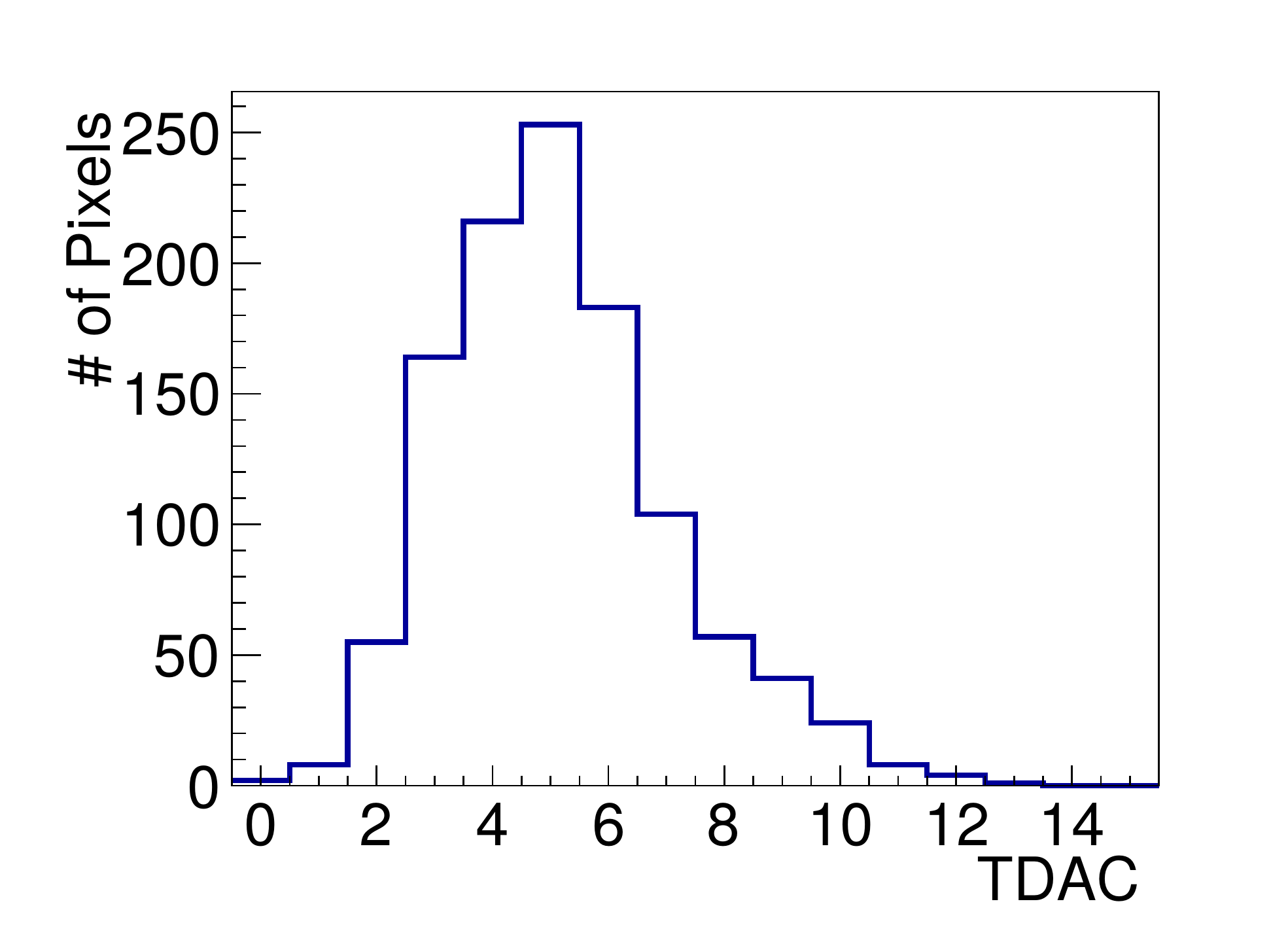}\hspace*{-0.3cm}
    \includegraphics[width=.5\columnwidth,keepaspectratio]{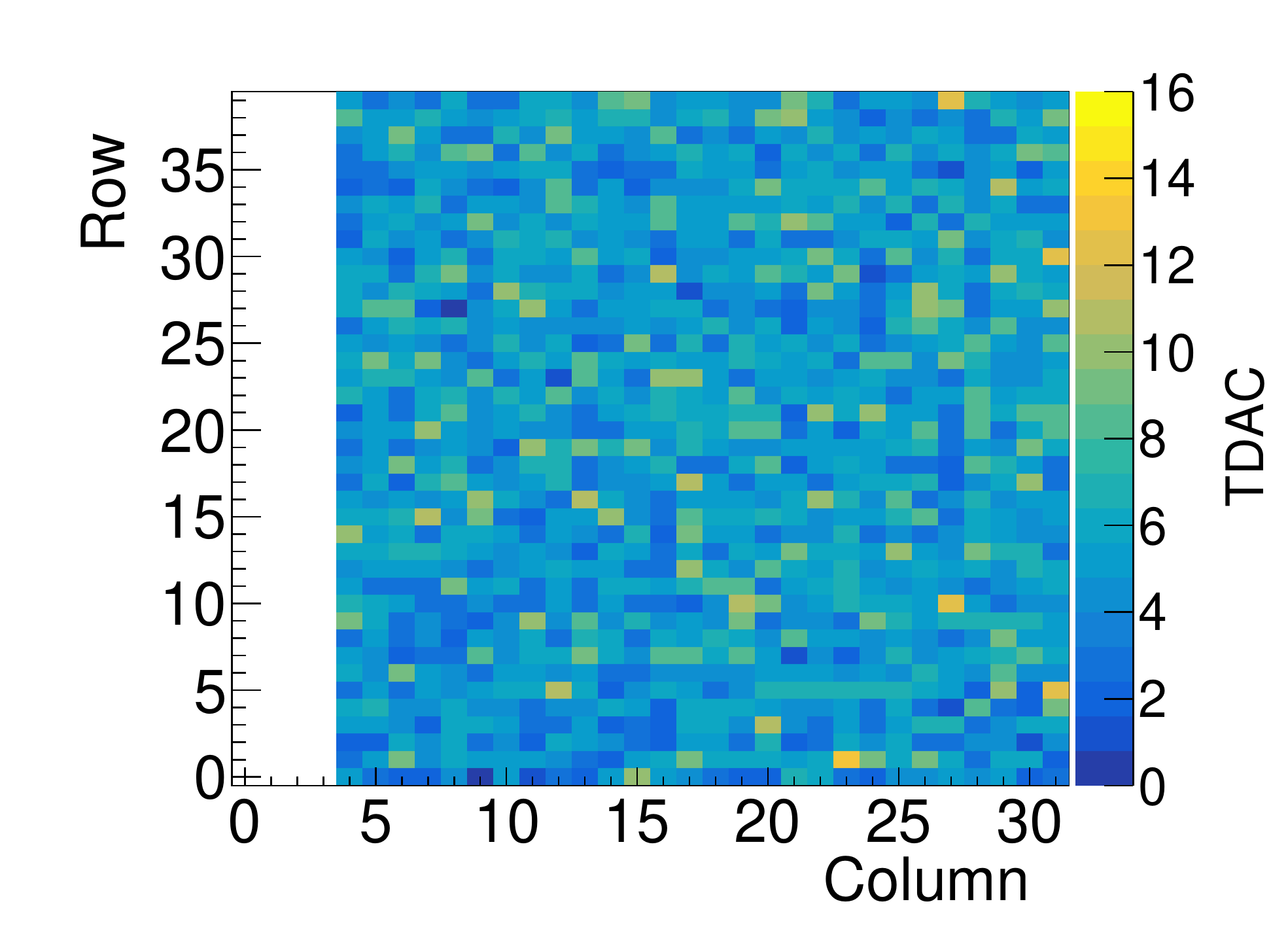}
    \caption{\label{fig:ntune}
\centering \SI{1e15}{\Neq} neutron irradiated (ID N115)}
    \end{subfigure}
    \begin{subfigure}[t]{\columnwidth}
    \includegraphics[width=.5\columnwidth,keepaspectratio]{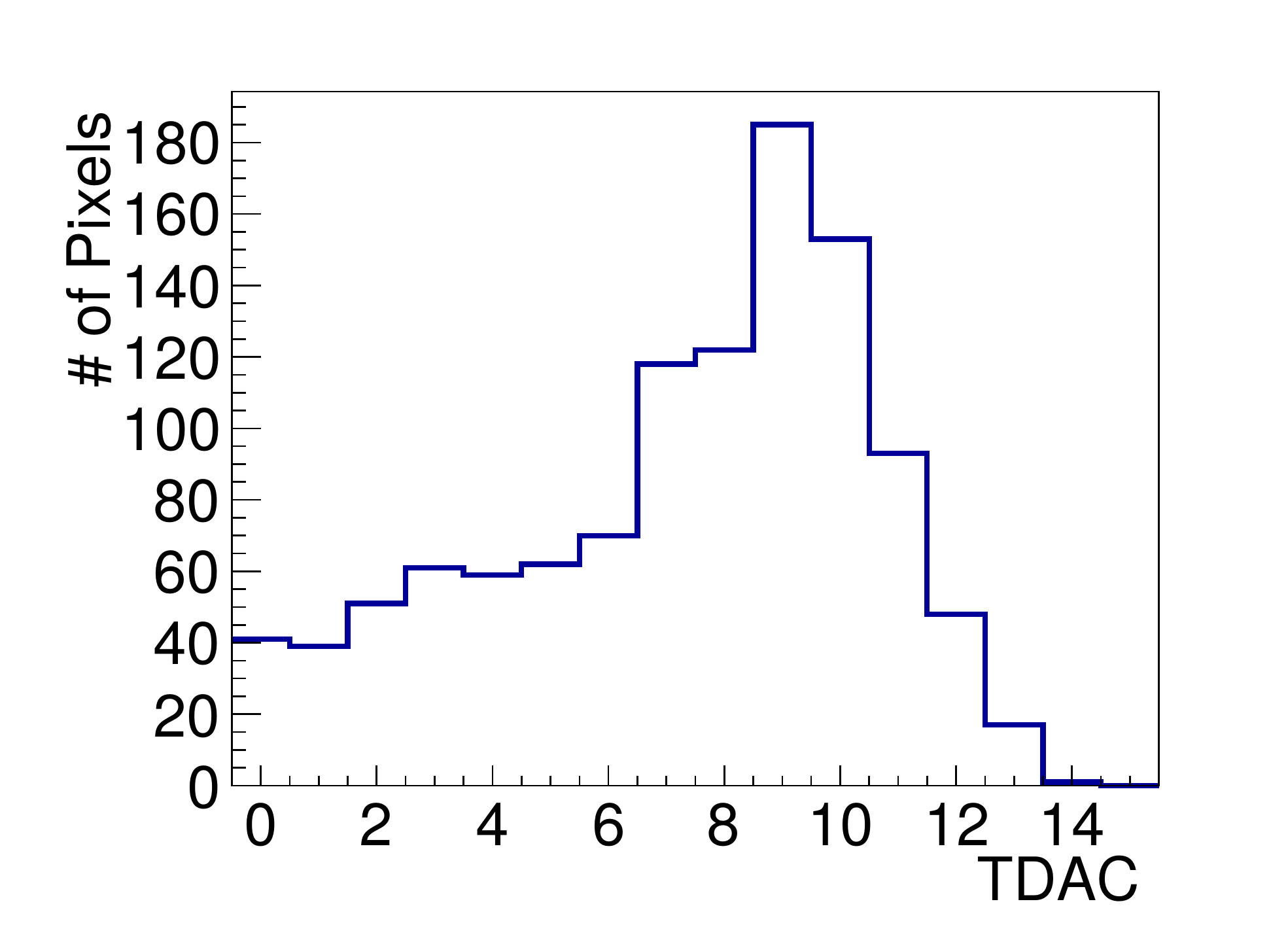}\hspace*{-0.3cm}
    \includegraphics[width=.5\columnwidth,keepaspectratio]{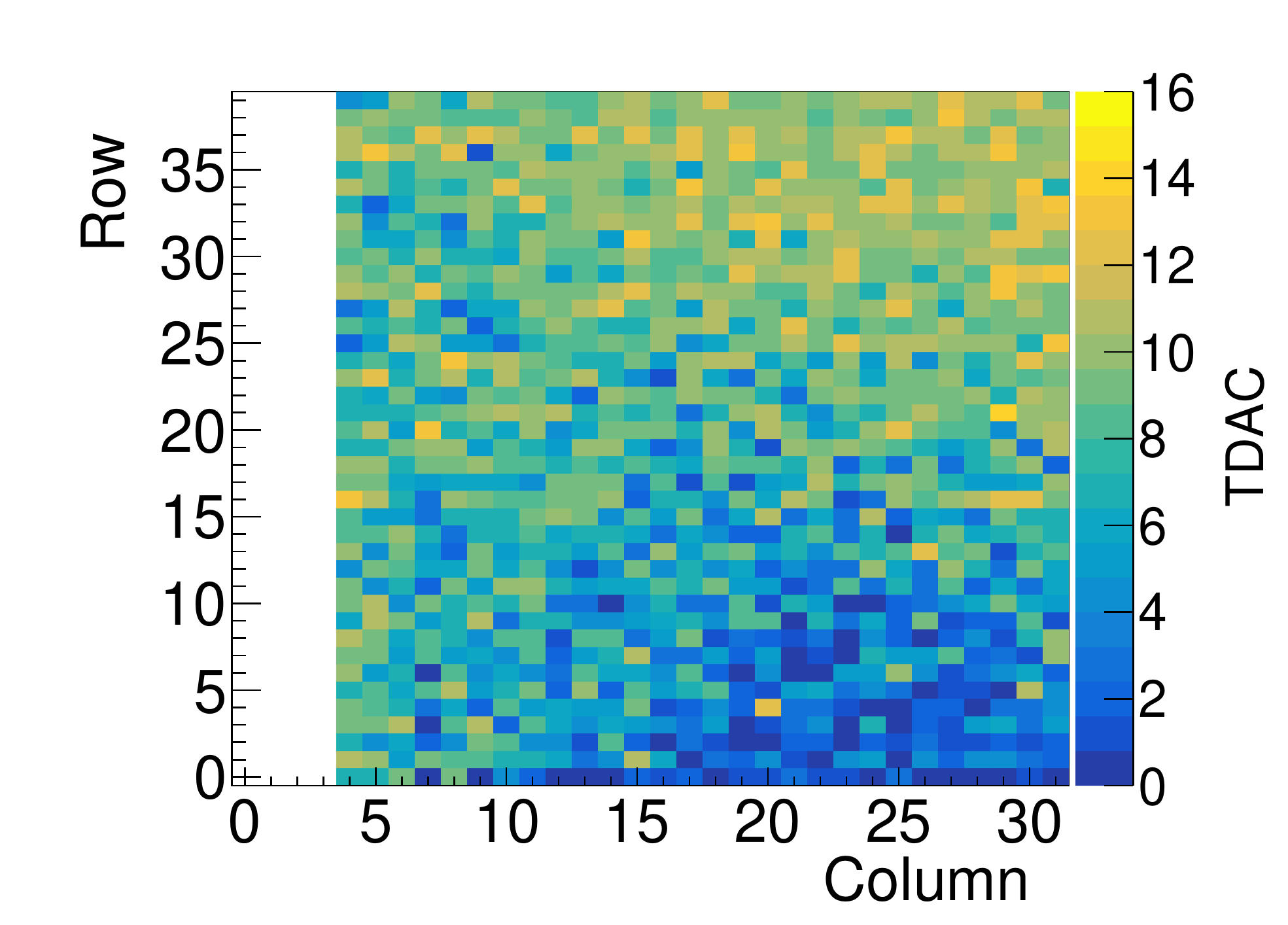}
    \caption{ \label{fig:ptune} \centering \SI{1.5e15}{p \per cm\squared} irradiated (ID~P1515)}
    \end{subfigure}
    \caption{\label{fig:tuning}\textbf{Left:} Pixel TDAC values for a
      non-irradiated, a neutron-irradiated and a proton-irradiated sensor.
      \textbf{Right:} Corresponding TDAC value shown as pixel matrix.}
    \end{figure}
The inhomogeneous proton irradiation prevented us from optimally tuning the sensors, thus compromising the following characterization studies.

\subsection{Test beam results}
Two MuPix telescopes \citep{TWEPP2017} are used as tracking reference for a test beam performed at the $\pi$M1 beam line at PSI in November 2016. 
One telescope  is used to characterize the proton irradiated MuPix, see figure \ref{fig:cooling}.
The other telescope is used to characterize the neutron irradiated MuPix.

The $\pi$M1 beam consists of a mixture of $\pi^+$, $e^+$, $\mu^+$ and protons, with $\pi^+$ being the dominant beam component.
The momentum is set to \SI{365}{MeV\per c} to select minimum ionizing $\pi^+$, which are expected to produce  1200 primary electrons in the depletion zone of a non-irradiated MuPix7 sensor at a depletion voltage of \SI{-85}{V}.
The particle rate of the beam is set to about \SI{100}{kHz} of which only about \SI{10}{kHz} are reconstructed by the telescopes due to limited acceptance.
Both MuPix telescopes are mechanically aligned with a precision of better than \SI{250}{\micro\meter} relative to each other.
A software alignment procedure is applied to correct for residual offsets with a precision of $\pm$~\SI{10}{\micro\meter}.

\subsection*{Efficiency and noise study} 
To study efficiency and noise, reference tracks are extrapolated to the DUT. On the DUT, clustering and crosstalk removal is applied.
A search window of \SI{800}{\micro\meter} radius and a time window of $\pm$~\SI{64}{ns} around the extrapolated track intersection is used to match hit clusters. 
The hit finding efficiency is defined as the number of matched tracks divided by the total number of extrapolated tracks and corrected for random coincidences.
The quoted efficiencies include all components of the readout system: hit
digitization, on-chip readout state machine, data transmission over the serial
link and front-end processing on the readout FPGA, i.e. time stamp sorting and
merging of data from the four telescope layers.

Figure \ref{fig:eff_example} shows exemplary results for the efficiency and noise of the tested samples as a function of the applied threshold reference. 
The noise consists of all unmatched clusters and is corrected for small
inefficiencies of the reference planes.
Due to the small size of the sensors in the reference planes (about 1/10 of
the beam profile) and large angle scatterers a significant fraction of beam particles entering the DUT
cannot be reconstructed.
This adds a constant noise floor of about \SI{20}{Hz/pixel} (\SI{7}{Hz/pixel}) for the proton (neutron) irradiated sensors.

The noise of the \SI{5e14}{\Neq} neutron irradiated sensor is similar to the non-irradiated
sensor: up to a reference threshold of about \SI{720}{mV} the noise stays below
\SI{10}{Hz\per pixel}; for higher threshold references (lower thresholds)
the noise increases exponentially.

The noise of the \SI{1.5e15}{protons\per cm\squared} proton irradiated sensor increases more rapidly
and already starts at low reference threshold  (high thresholds).
Proton and neutron irradiated sensors with a depletion voltage of \SI{-60}{V}
reach efficiencies of  about \SI{95}{\percent}, similar to the non-irradiated
sensor, however, at the expense of factor 10-100 higher noise levels
and increased leakage currents.
Higher efficiencies can be reached by further increasing the depletion-voltage.

 \begin{figure}[tbp]
 \centering
 \begin{subfigure}[t]{.9\columnwidth}
	 	 \centering
	 	 \includegraphics[width=.97\columnwidth,keepaspectratio]{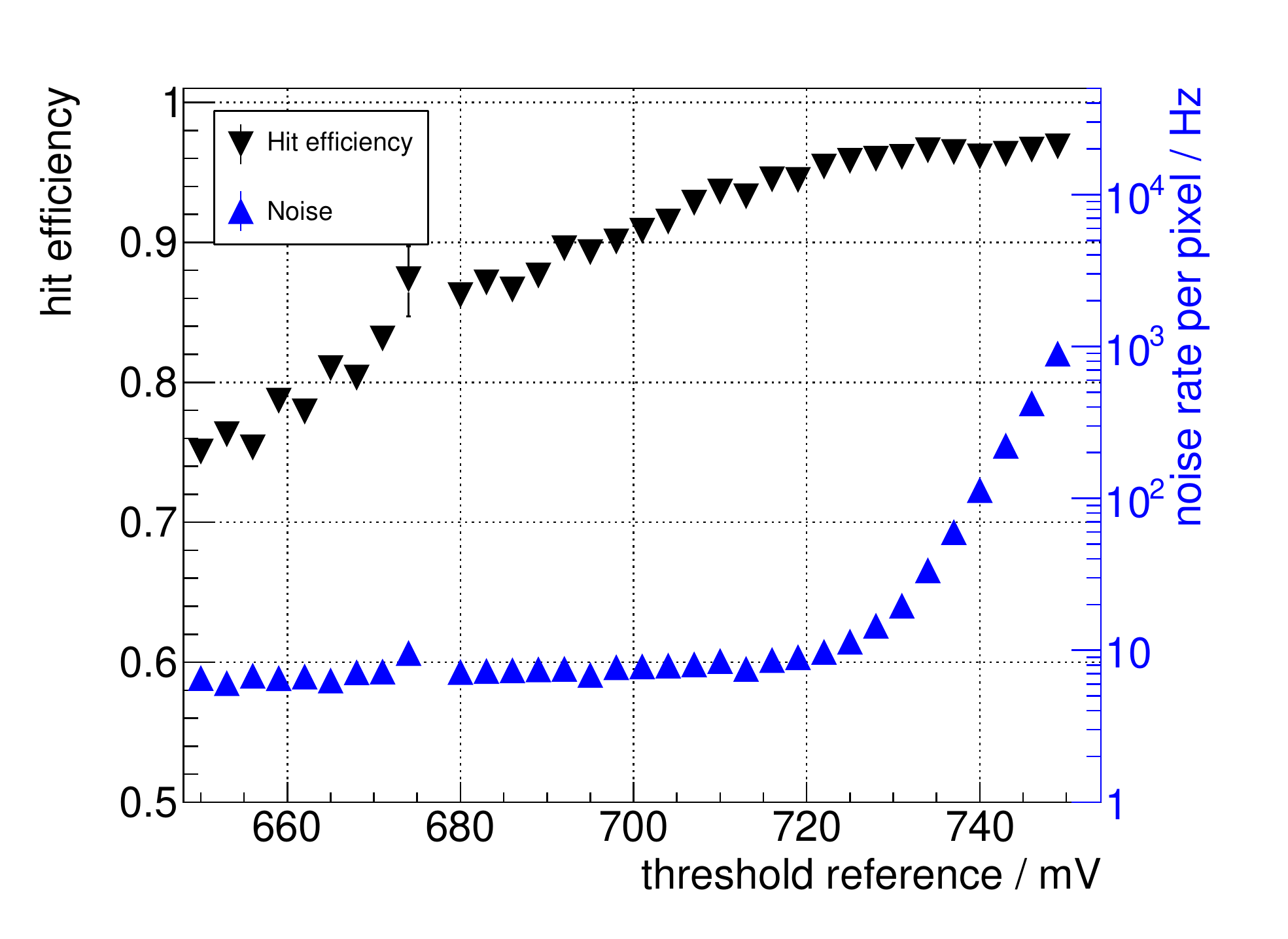}
 	\caption{\label{fig:eff_non} non-irradiated}
 \end{subfigure}
\begin{subfigure}[t]{.9\columnwidth}
		 \centering
		 \includegraphics[width=.97\columnwidth,keepaspectratio]{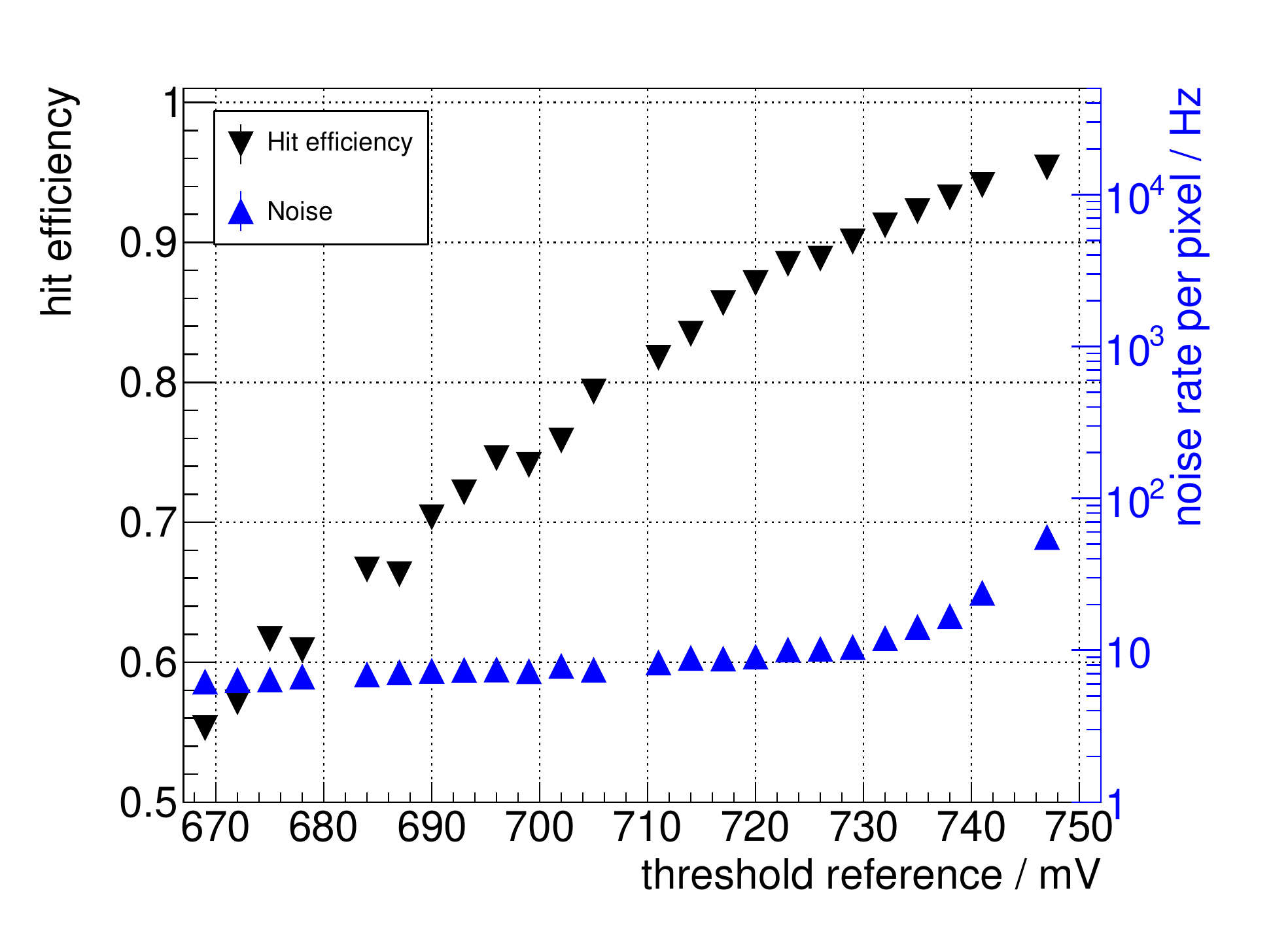}
		\caption{\label{fig:eff_n} \SI{5e14}{\Neq}}
	\end{subfigure}
\begin{subfigure}[t]{.9\columnwidth}
		 \centering
		 \includegraphics[width=.97\columnwidth,keepaspectratio]{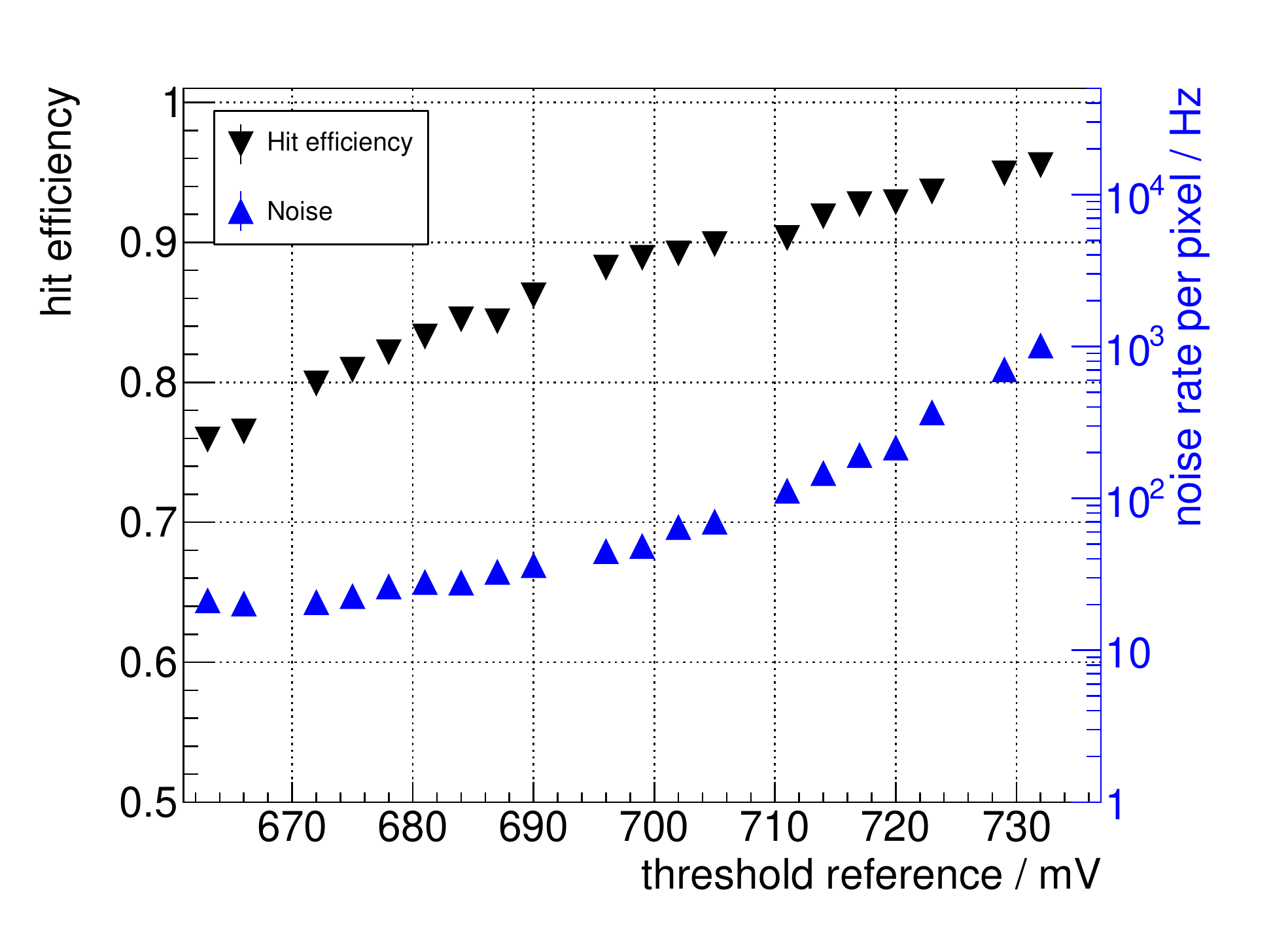}
		\caption{\label{fig:eff_p} \SI{1.5e15}{p\per\centi \meter\squared}}
	\end{subfigure}
\caption{\label{fig:eff_example} Efficiency and noise as function of the threshold reference for a depletion voltage of \SI{-60}{V} at T$_{\text{MuPix}} \approx$~\SI{8}{\celsius} for non-, neutron- and proton irradiated sensors.}
\end{figure}

The influence of the depletion voltage on the efficiency and noise as a
function of the threshold reference is shown in figure~\ref{fig:hv_dep} for a
neutron irradiated sensor with \SI{5e15}{\Neq} which was operated
at a sensor temperature of about \SI[parse-numbers = false]{8}{\celsius}. 
The sensor efficiency increases with the applied depletion voltage, consistent with the expectation that the active depletion zone grows proportional to $\sqrt{\text{U}_\text{HV}}$, leading to higher signals.
At very high negative voltages additional avalanche effects contribute to charge
amplification, which sets in at about \SI{-80}{V} for non-irradiated sensors
and shifts to slightly higher negative voltages for irradiated sensors
\cite{ChargeCollection}.
For a depletion voltage of \SI{-85}{V} an efficiency of about
\SI{90}{\percent} is measured at a threshold reference of \SI{715}{mV} and a noise rate of about \SI{100}{Hz} per pixel.
The noise significantly increases for larger threshold references.
The shift of the threshold curves for the different depletion voltages can be
explained by tuning effects:
different VPDAC values and tune thresholds are used for the  \SI{-70}{V} and  \SI{-85}{V} measurements compared to the \SI{-40}{V} and \SI{-60}{V} measurements, as seen in table~\ref{tab:tunings}.
This results in a shift of the effective threshold by about \SI{15}{mV}.

 \begin{figure}[tb]
 \centering
\begin{subfigure}[t]{0.95\columnwidth}
\includegraphics[width=\columnwidth,keepaspectratio]{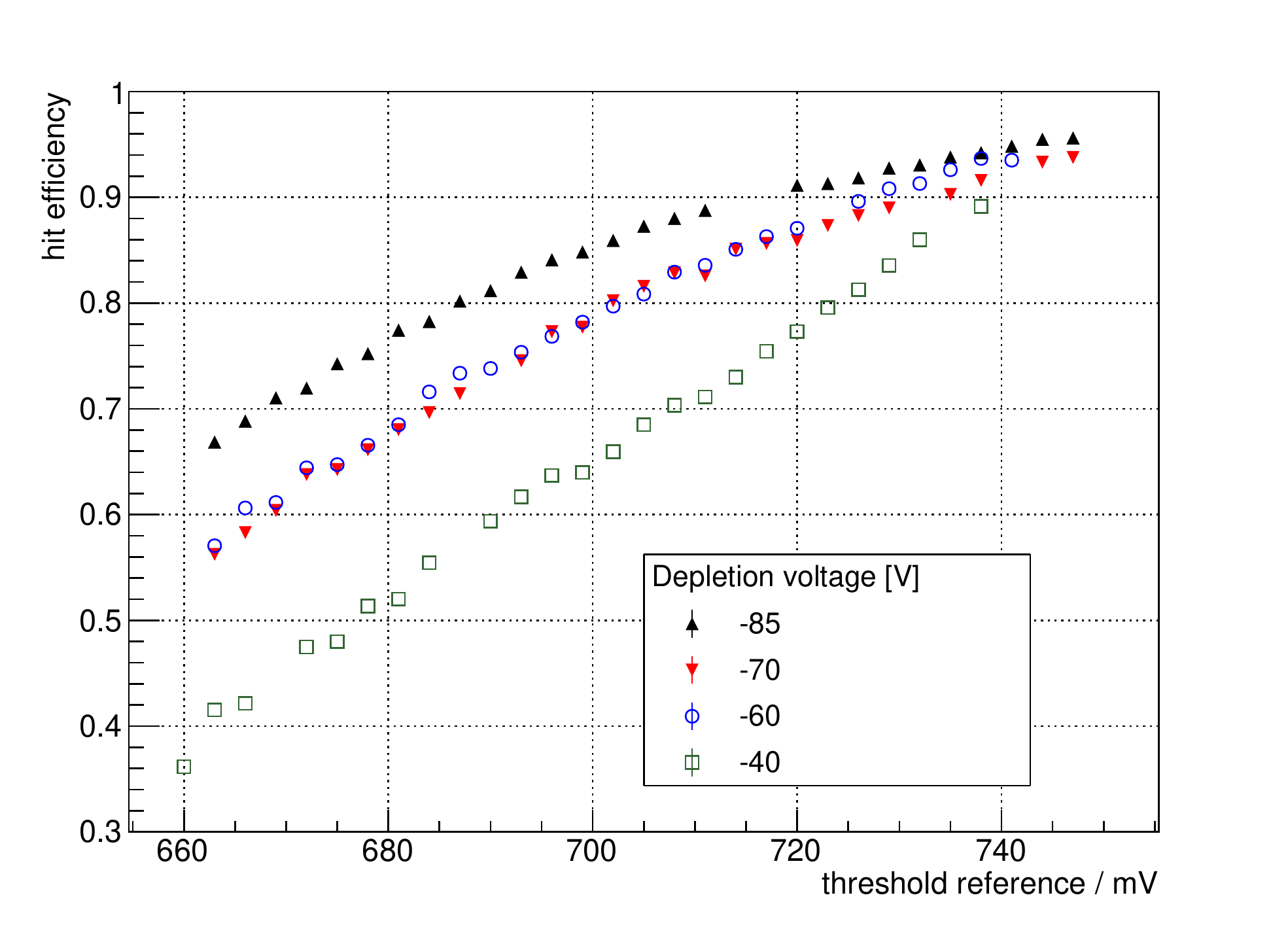}
		\caption{\label{fig:hv_eff} Efficiency}
	\end{subfigure}
\begin{subfigure}[t]{0.95\columnwidth}
			\includegraphics[width=\columnwidth,keepaspectratio]{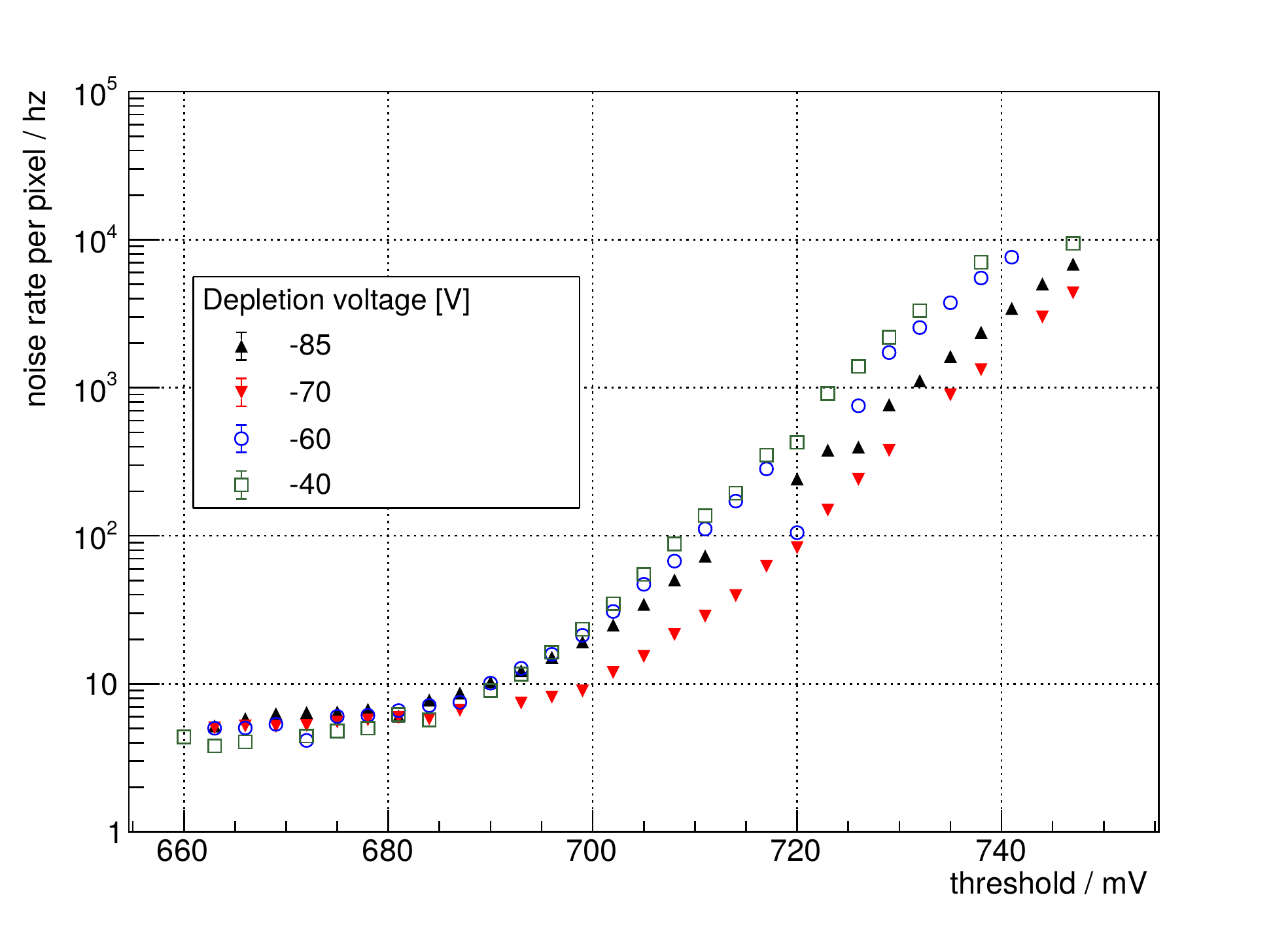}
		\caption{\label{fig:hv_noise} Noise rate}
	\end{subfigure}
\caption{\label{fig:hv_dep}Efficiency and noise rate as a function of the
  threshold reference for the \SI{5e15}{\Neq} neutron-irradiated MuPix7
  sensor. Results are shown for different HV settings at
  T$_{\text{MuPix}}$~$\approx$~\SI{8}{\degreeCelsius}.
 Note that the sensor was configured at different tune thresholds and
 different tuning strength (VPDAC) for the different depletion voltages, see table~\ref{tab:tunings} leading to
 threshold curve shifts of about \SI{15}{mV}.
}
\end{figure}
\interfootnotelinepenalty=10000

For the different irradiated samples the efficiency and noise measurements are summarized
in figure \ref{fig:comparison} as a function of threshold value.
Data samples with a common, but reduced, HV of \SI{-60}{V} are chosen here to
allow for a systematic comparison of all irradiated sensors\footnote{Sensor N514 was accidentally damaged after taking
  data at \SI{-60}{V}.}.
Small differences in the efficiency are expected from the different telescope geometries: The neutron irradiated samples are glued on a carrier and placed behind the three reference layers. They have by \SI{0.3}{\percent} to \SI{0.8}{\percent} reduced efficiencies due to undetected particle losses with large angle scattering in the third layer.

In general, neutron and proton irradiated samples show similar
performance, considering the threshold variations due to different VPDAC settings and
the limited threshold scan for the neutron-irradiated sensor which does not
reach very low effective thresholds.
Accounting for these differences the data show
overall an efficiency loss and noise increase  for increasing neutron and
proton fluences.
Although the MuPix irradiated to \SI{1.5e15}{p \per cm\squared} reaches almost
similar efficiencies than the non-irradiated one, it shows a significant
noise increase.
At very low threshold reference values (high threshold) a beam induced noise floor of \SI{20}{Hz} 
and \SI{7}{Hz} is measured for the proton and neutron irradiated sensors, respectively.
 \begin{figure}[tb]
 \centering
\begin{subfigure}[t]{0.95\columnwidth}
\includegraphics[width=\columnwidth,keepaspectratio]{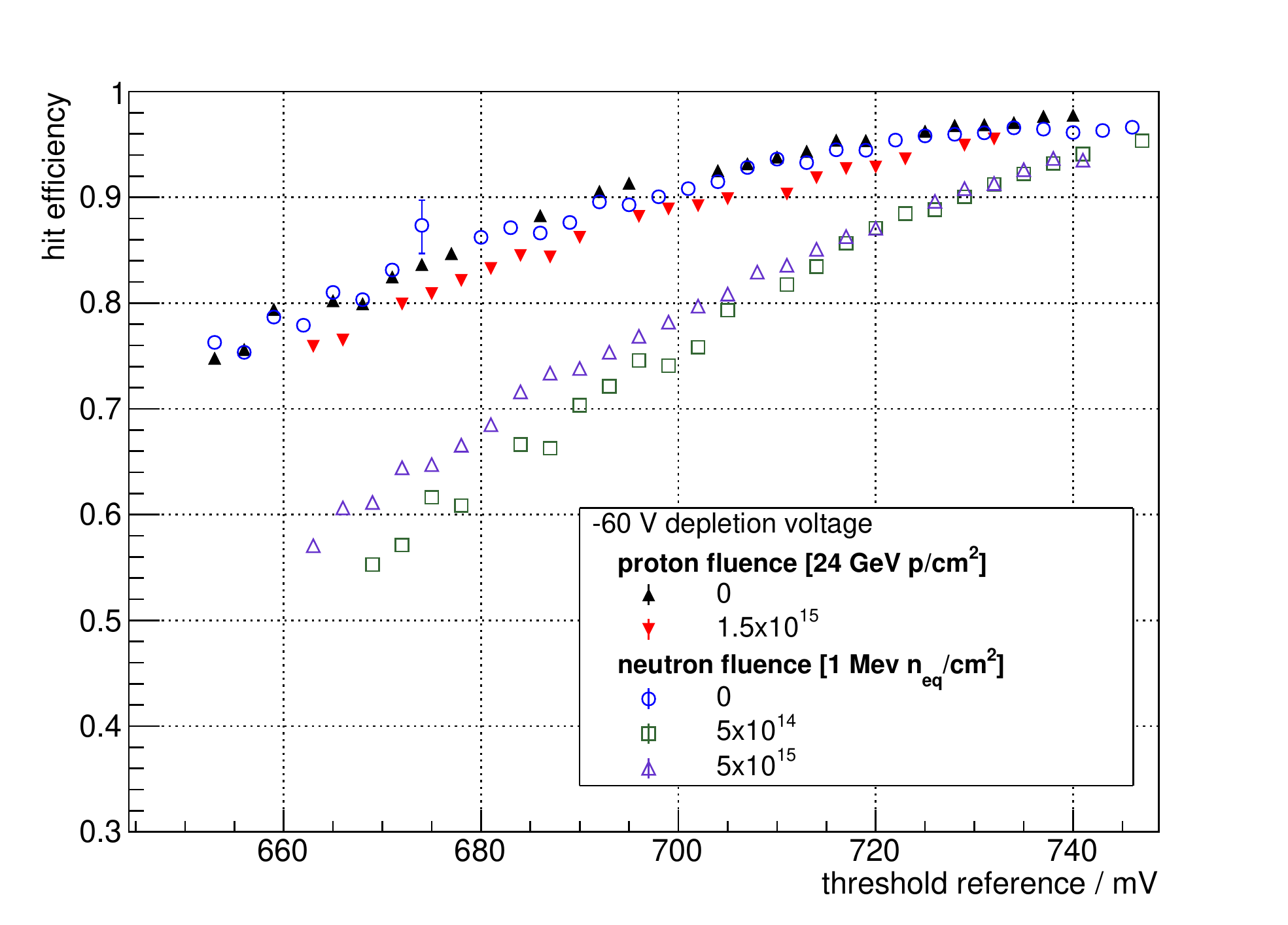}
		\caption{\label{fig:comparison_eff} Efficiency}
	\end{subfigure}
\begin{subfigure}[t]{0.95\columnwidth}
			\includegraphics[width=\columnwidth,keepaspectratio]{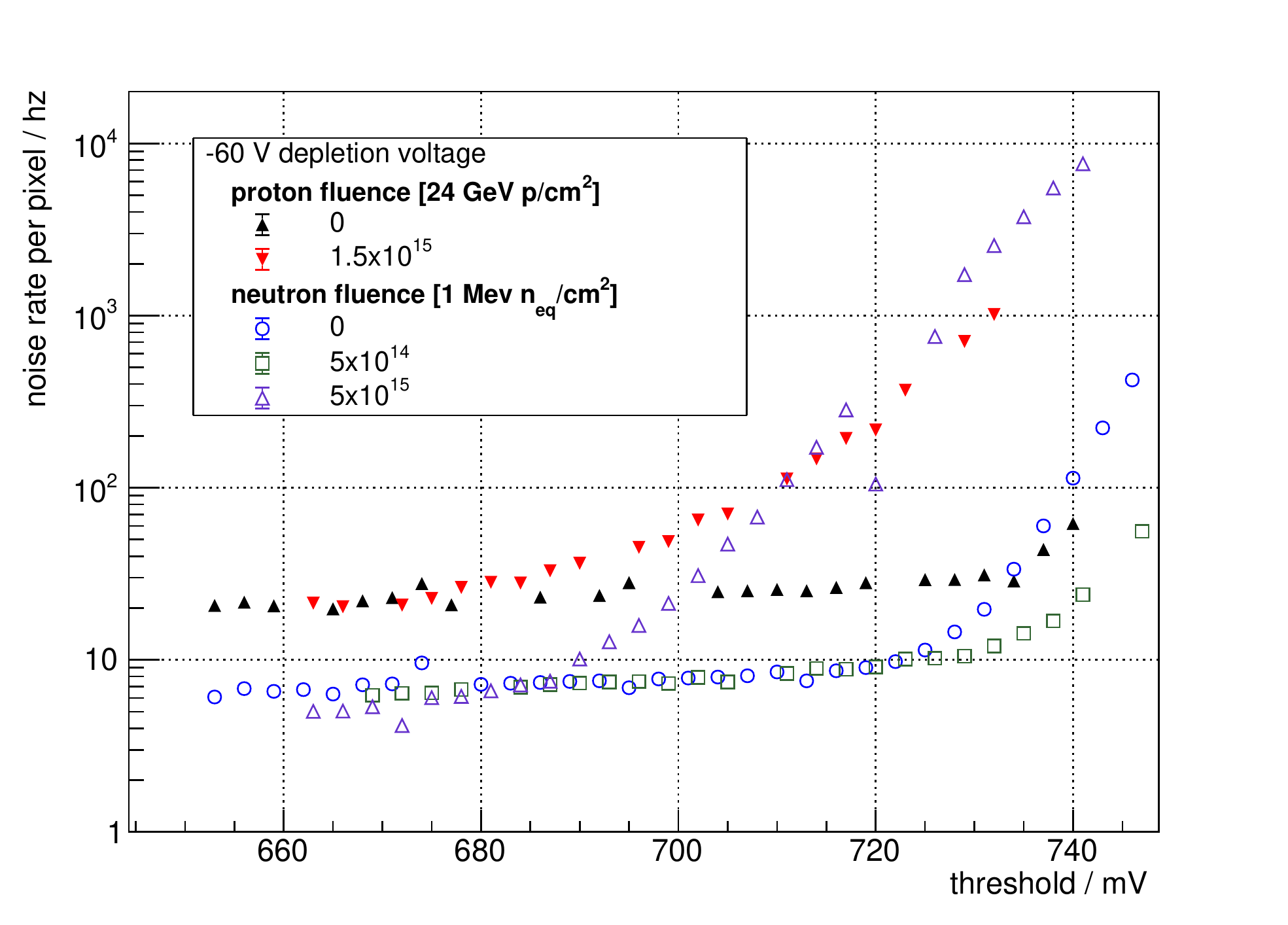}
		\caption{\label{fig:comparison_noise} Noise rate}
	\end{subfigure}
\caption{\label{fig:comparison}Efficiency and noise rate as a function of the
  threshold reference for all tested samples for HV = \SI{-60}{V} and
  T$_{\text{MuPix}}$~$\approx$~\SI{8}{\degreeCelsius}. 60~V is chosen, as one
  sensor was damaged after taking data at 60~V.
Note that the sensors were configured at different tune thresholds and with different tuning strength (VPDAC), see table~\ref{tab:tunings}), leading to threshodl curve shifts of about \SI{50}{mV}.}
\end{figure}

The automated threshold tuning procedure leads to strong correlations between
the efficiency and noise measurements and the threshold used for tuning,
also visible as shifts of the threshold curves in figures~\ref{fig:hv_dep}
and~\ref{fig:comparison}.
To compare the different samples in a more setting independent way, the
efficiencies are determined for fixed average  noise rates of \SI{40}{Hz} (\SI{200}{Hz}) per pixel\footnote{For LHC experiments noise occupancies are typically required to be
  below \SI{1e-6}{}, corresponding to \SI{40}{Hz} noise per pixel at \SI{40}{MHz} bunch crossing frequency.}.
The noise floor of \SI{7}{Hz} (\SI{20}{Hz}) for the neutron (proton) irradiated samples is subtracted.
The resulting efficiencies are shown in figure \ref{fig:summary} for depletion
voltages of \SI{-60}{V} and \SI{-85}{V}.
The efficiencies of the proton- and neutron-irradiated sensors show a moderate
reduction up to a fluence of \SI{1.5e15}{\per cm\squared}. 
For higher fluences $>$~\SI{1.5e15}{\per cm\squared} the efficiency decreases
more strongly and falls below 90~\%. 
The efficiency difference between the average per pixel noise rates of \SI{40}{Hz} and
\SI{200}{Hz} indicates the possible efficiency gain if the sensors are cooled
to even lower temperatures.

 \begin{figure}[tbp]
 	\centering
  		\includegraphics[width=\columnwidth,keepaspectratio]{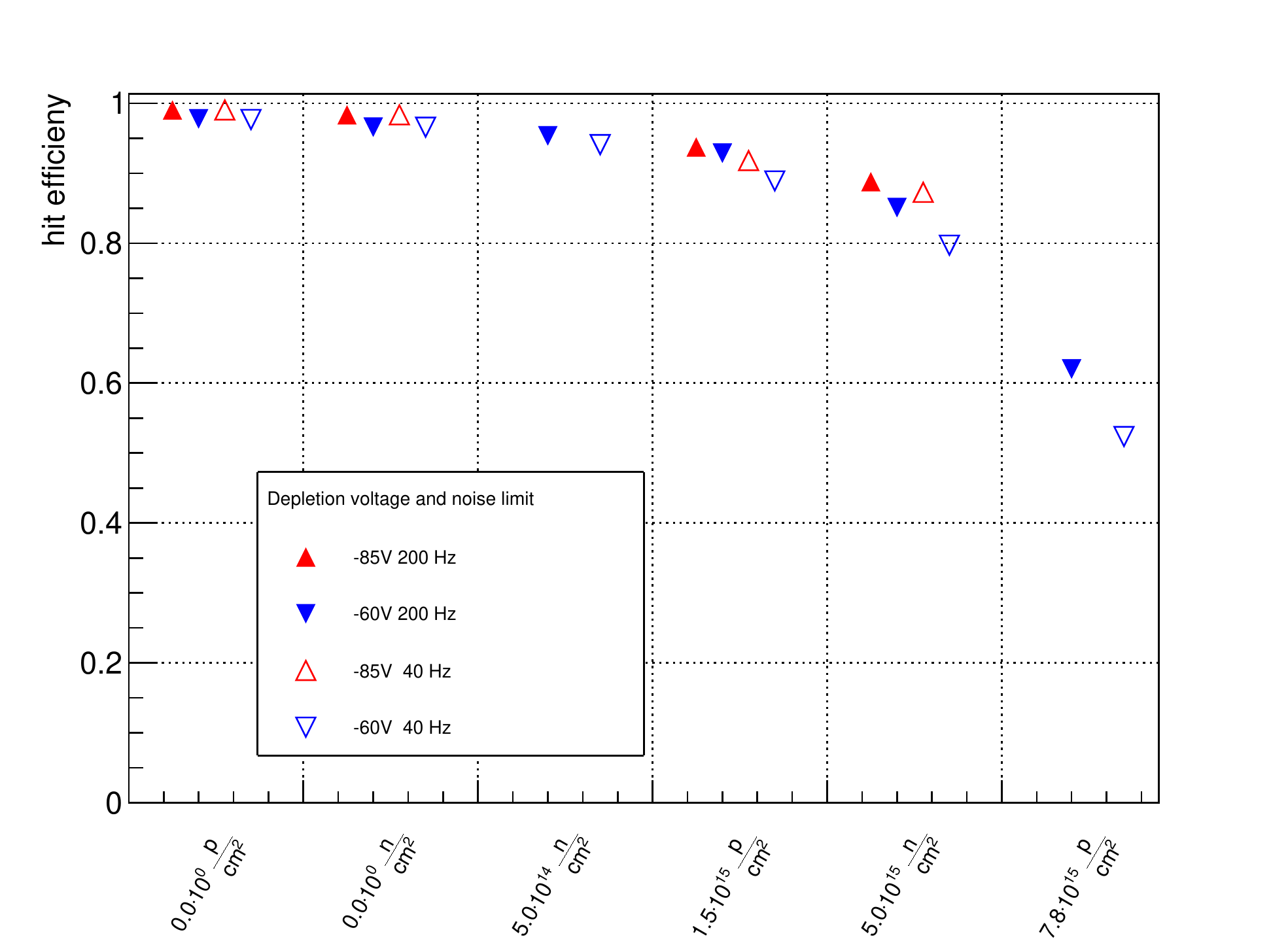}
 \caption{\label{fig:summary} Efficiencies as function of the fluence for
   depletion voltages of \SI{-60}{V} ({\sl blue} symbols) and \SI{-85}{V} ({\sl red}
   symbols) and  T$_{\text{MuPix}}$~$\approx$~\SI{8}{\degreeCelsius}. The filled triangles show the results for an average per pixel noise rate of \SI{200}{Hz} and the empty ones for \SI{40}{Hz}. The beam induced noise floor is subtracted. Statistical error bars are too small to be seen.}
 \end{figure}
 \noindent

\subsection*{Time resolution}

The time resolution of the MuPix7 DUT is measured relative to the averaged time stamps of the hits from reference tracks.
A Gaussian fit is applied to a histogram of the time differences. 
The standard deviation, $\sigma$, of the fit defines the time resolution, which is corrected for the limited resolution of the reference sensors
\begin{equation}
	{\sigma^2}_{DUT} = {\sigma^2}_{Fit} - {\sigma^2}_{Ref}
\end{equation}
assuming a time resolution of $\sigma_{MuPix}=\SI{14.2}{ns}$ for the non-irradiated
MuPix7~\cite{mp7,mp7timing}. Using ${\sigma}_{Ref}= \frac{1}{3}\sqrt{3 \cdot
  {\sigma^2}_{MuPix}}$ the time resolution of the DUT is measured for all
MuPix and for different depletion voltages, see figure~\ref{fig:timing}.
The time resolution of the non-irradiated MuPix of about \SI{14}{ns} is in agreement with previous measurements \cite{mp7}.
It stays constant for fluences up to \SI{1.5e15}{p\per cm\squared} and is also
rather independent of the threshold setting and therefore the noise limit.
For higher proton and neutron fluences, the time resolution of the sensor becomes significantly worse.

 \begin{figure}[tbp]
 \centering
			\includegraphics[width=\columnwidth,keepaspectratio]{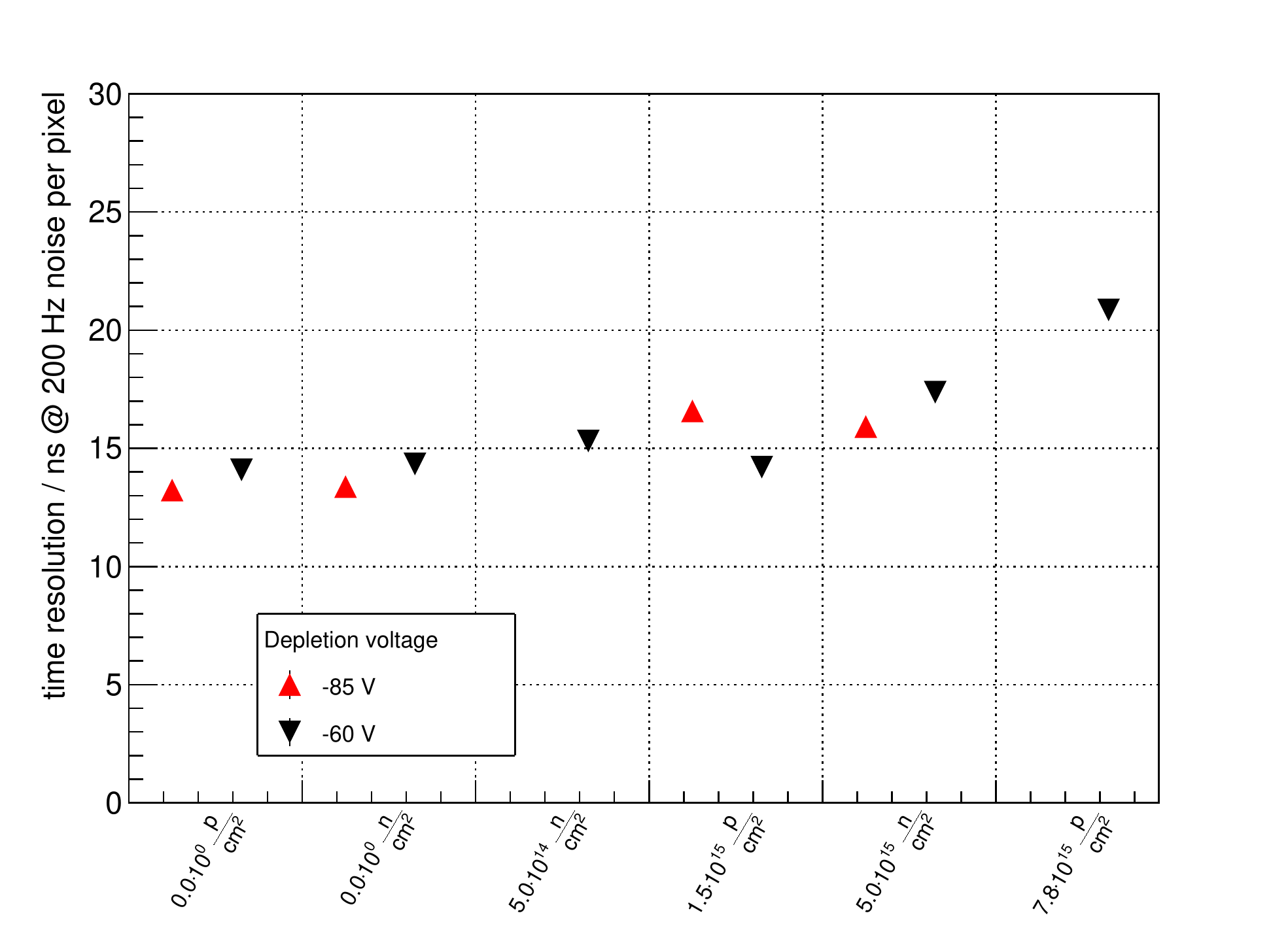}
\caption{\label{fig:timing} Time resolution expressed in Gaussian  $\sigma$ as function of the irradiation dose for depletion voltages of \SI{-60}{V} and \SI{-85}{V} at an exemplary average noise rate of 200 Hz and T$_{\text{MuPix}}$~$\approx$~\SI{8}{\degreeCelsius}. Statistical error bars are too small to be seen.}
\end{figure}

\section{Conclusion}
Proton and neutron irradiated samples of HV-MAPS prototypes with a particle fluence of up to \SI{7.8e15}{p\per cm\squared} were tested in the laboratory and at a PSI test beam. 
All sensors are fully functional after one year of annealing at room temperature.

The MuPix7 samples have a non-radiation-hard design and were realized on a  \SI[parse-numbers=false]{10-20}{\ohm cm} substrate with a depletion zone of \SI[parse-numbers=false]{10-14}{\micro m} at a depletion voltage \SI{-85}{V} before irradiation. 
For the irradiated samples increased noise rates and leakage currents in the
pixel matrix are observed. The proton irradiated samples show significantly higher leakage currents as the neutron irradiated sensors for similar particle fluences in the laboratory.

At PSI, efficiency and noise studies with sensors cooled down to \SI{8}{\celsius} have been carried out.
At an optimal depletion voltage of \SI{-85}{V} and
at a maximum noise limit of \SI{40}{Hz \per pixel}
efficiencies \SI{\ge 90}{\percent} are measured for all sensors with a dose of
up to \SI{1.5e15}{p\per cm\squared}. For higher proton and neutron fluences a
significant performance degradation is observed. 
The time resolution for all irradiated sensors is below \SI{22}{ns}, compared to about \SI[parse-numbers=false]{14}{ns} time resolution of the non-irradiated references.
Only a small time resolution decrease for fluences of up to \SI{5e15}{\Neq} is
observed. 
During full operation in the test beam the overall noise seems to be strongly influenced by bulk damage induced charge trapping.

Despite the non-radiation-hard design and the very small depletion zone of the standard AMS H18 
process, the MuPix7 shows high radiation tolerance emphasizing the potential of the AMS H18 process for usage in harsh radiation environments. 
The radiation tolerance of the synthesized and fast readout state machine
logic is demonstrated for the first time for a fully monolithic prototype
realized in HV-CMOS technology. 

\section*{Acknowledgments}
N.~Berger and D.~vom~Bruch thank the \textit{Deutsche Forschungsgemeinschaft} for supporting them and the Mu3e project through an Emmy Noether grant. 
S.~Dittmeier and L.~Huth acknowledge support by the IMPRS-PTFS. 
A.-K.~Perrevoort acknowledges support by the Particle Physics beyond the
Standard Model research training group [GRK 1940].
H.~Augustin and A.~Herkert acknowledge support by the HighRR research training group [GRK
  2058]. 
N.~Berger thanks  the PRISMA Cluster of Excellence for
support. 
This project was also supported by BMBF grant 05H15VHCA1.
We would like to thank PSI for valuable test beam time.
The authors gratefully acknowledge the support by the CERN PS and SPS instrumentation team
and would like to sincerely thank Prof. Dr. V. Cindro and the team of the
TRIGA reactor in Ljubljana for performing neutron irradiations for this
publication.  

\bibliographystyle{unsrt_collab_comma}
\bibliography{mybibfile.bib}

\end{document}